\documentclass[12pt]{article}
\usepackage{amsmath,epsfig,graphicx,verbatim}
\usepackage{epstopdf}
\usepackage{color}
\usepackage{subfigure}

\include{epsf} 
\def\ltap{\raisebox{-.6ex}{\rlap{$\,\sim\,$}} \raisebox{.4ex}{$\,<\,$}} 
\def\gtap{\raisebox{-.6ex}{\rlap{$\,\sim\,$}} \raisebox{.4ex}{$\,>\,$}}

\newcommand\as{\alpha_{\mathrm{S}}} 
 
\def\beq{\begin{equation}} 
\def\eeq{\end{equation}} 
\def\beeq{\begin{eqnarray}} 
\def\eeeq{\end{eqnarray}} 
\def\bom#1{{\mbox{\boldmath $#1$}}} 
\def\to{\rightarrow}

\def\qt{q_T}
\def\bqt{\bom{q_T}}

\def\qh{\bom{{\hat q}_T}}
\def\kh{\bom{{\hat k}_T}}

\begin{document} 
\begin{titlepage}
\renewcommand{\thefootnote}{\fnsymbol{footnote}}
\begin{flushright}
     ZU-TH 04/17
     \end{flushright}
\par \vspace{10mm}

\begin{center}
{\Large \bf Azimuthal asymmetries in QCD hard scattering:
\\[0.2cm]
 infrared safe but divergent}
\end{center}

\par \vspace{2mm}
\begin{center}
{\bf Stefano Catani${}^{(a)},$ Massimiliano Grazzini${}^{(b)}$}
and {\bf Hayk Sargsyan$^{(b)}$\\
}

\vspace{5mm}

$^{(a)}$ INFN, Sezione di Firenze and Dipartimento di Fisica e Astronomia,\\ 
Universit\`a di Firenze,
I-50019 Sesto Fiorentino, Florence, Italy

$^{(b)}$ Physik-Institut, Universit\"at Z\"urich, 
CH-8057 Z\"urich, Switzerland

\vspace{5mm}

\end{center}

\par \vspace{2mm}
\begin{center} {\large \bf Abstract} \end{center}
\begin{quote}
\pretolerance 10000

We consider high-mass systems of two or more particles that are produced by QCD
hard scattering in hadronic collisions. 
We examine the azimuthal correlations between
the system and one of its particles.
We point out
that the perturbative QCD computation of such
azimuthal correlations and asymmetries can lead to divergent results at fixed
perturbative orders.
The fixed-order divergences affect basic (and infrared safe) quantities 
such as the total cross section at fixed (and arbitrary) values of the
azimuthal-correlation angle $\varphi$.
Examples of processes with fixed-order divergences are heavy-quark pair
production, associated production of vector bosons and jets, dijet and diboson
production. 
A noticeable exception is the production of high-mass lepton pairs through
the Drell--Yan mechanism of quark-antiquark annihilation.
However, even in the Drell--Yan process, fixed-order divergences arise in the
computation of QED radiative corrections.
We specify general conditions that produce the divergences by discussing
their physical origin in fixed-order computations.
We show lowest-order illustrative results for $\cos(n\varphi)$ asymmetries (with $n=1,2,4,6$) in top-quark pair
production and associated production of a vector boson and a jet
at the LHC.
The divergences are removed by a proper all-order resummation procedure of the
perturbative contributions. Resummation leads to azimuthal asymmetries that are
finite and computable. 
We present first quantitative results of such a resummed computation for the $\cos(2\varphi)$ asymmetry in top-quark pair
production at the LHC.

\end{quote}

\vspace*{\fill}
\begin{flushleft}
March 2017

\end{flushleft}
\end{titlepage}

\setcounter{footnote}{1}
\renewcommand{\thefootnote}{\fnsymbol{footnote}}
\section{Introduction}
\label{sec:intro}

Angular distributions of final-state particles produced by high-energy collisions
are known to be relevant observables for the understanding of the 
underlying dynamics of their production mechanism.
In this paper we deal with azimuthal-angle distributions and related asymmetries.

In the case of collisions that are produced by spin polarized particles,
azimuthal angles with respect to the spin direction can be specified and measured:
the analysis of the ensuing azimuthal correlations is a much studied topic
and the subject of intense ongoing investigations (see, e.g., 
Ref.~\cite{Aschenauer:2015ndk} and references therein).

Azimuthal correlations in spin unpolarized collisions are less studied.
A relevant exception is the lepton azimuthal distribution 
\cite{Collins:1977iv} of high invariant mass lepton pairs
that are produced in hadron collisions through the Drell--Yan (DY) mechanism
of quark-antiquark annihilation.
Angular distributions for the DY process are a well known topic that has been
deeply studied at both the theoretical and experimental levels
(see, e.g., Refs.~\cite{Khachatryan:2015paa, Aad:2016izn, Lambertsen:2016wgj}
and references therein).

In the case of spin unpolarized collisions, much attention to azimuthal
asymmetries has been devoted in recent years within the context of a QCD framework
based on the 
introduction of non-perturbative transverse-momentum dependent (TMD)
distributions of quarks and gluons inside the colliding unpolarized hadrons.
In particular, the TMD distribution of linearly-polarized gluons 
\cite{Mulders:2000sh}
inside unpolarized hadrons plays a distinctive role since it leads to specific
$\cos(2\varphi)$ and $\cos(4\varphi)$ modulations of the dependence on the
azimuthal-correlation angle $\varphi$. TMD distribution studies of such
modulations have been performed, for instance, for the production process 
of heavy quark-antiquark ($Q{\bar Q}$) pairs \cite{Pisano:2013cya}
and for the associated production of a virtual photon and a jet ($\gamma^* +
1$~jet) \cite{Boer:2017xpy}. Many more related studies can be found in the list of
references of Ref.~\cite{Boer:2017xpy}.

In this paper we consider spin unpolarized collisions and, specifically, we focus
our discussion on hadron--hadron collisions, although many features that we
discuss are generalizable to (and, hence, valid for) lepton--hadron and
lepton--lepton collisions.
We consider the production of a system $F$ of two or more particles with a high
value of the total invariant mass, and we examine the azimuthal correlations
between the momenta of the system and of one of its particles.
Roughly speaking, the relevant azimuthal angle $\varphi$ is related to the
difference between the azimuthal angles of the transverse momenta $\bqt$
of the system and $\bom{p_T}$ of the particle.
Considering high values of the invariant mass $M$ of the system $F$
(additional kinematical cuts 
on the momentum of the
produced particle can be applied or required), 
we select the hard-scattering regime of the production
process. In this regime, azimuthal correlations are infrared and collinear safe
observables \cite{Sterman:1977wj}, and they can be studied and computed by
applying the standard QCD framework of factorization 
in terms of perturbative partonic cross sections and parton distribution functions
(PDFs) of the colliding hadrons.
Starting from the results in
Refs.~\cite{Nadolsky:2007ba, Catani:2010pd} and, especially, from those in 
Ref.~\cite{Catani:2014qha} and the analysis on the role of soft wide-angle
radiation that is performed therein, 
we develop and present some general (process-independent)
considerations on azimuthal correlations and asymmetries.

We point out with some generality that, in spite of the infrared and collinear
safety of azimuthal-correlation observables, their computation at fixed
perturbative orders leads to divergences. 
The divergences appear also in basic quantities such as the total cross section
at fixed (and arbitrary) values of the azimuthal-correlation angle $\varphi$.
Example of processes with {\em fixed order} (f.o.) divergences are
heavy-quark pair ($Q{\bar Q}$) production, 
associated production of vector ($V$) or Higgs bosons and jets (e.g.,
`$V+1$\,jet'
with $V=Z,W^\pm,\gamma^*$), dijet and diboson
production. We support our theoretical discussion by performing the numerical
calculation of $\cos(n\varphi)$ modulations (for some values of $n$) at the lowest
perturbative order for some specific processes. The numerical results are found to
be consistent with the divergent behaviour if $n=2,4,6$ in the case of top-quark
pair ($t{\bar t}$) production and if $n=1,2,4,6$
in the case of associated `$Z+1$\,jet'
production.
A noticeable exception to the appearance of QCD divergences is the case of the DY
process.
However, even in the case of the DY process, we predict the presence of f.o.
divergences in the perturbative computation of QED radiative corrections.
Moreover, analogous divergences arise in the case of lepton--lepton collisions
from the computation of radiative corrections
in a pure QED context (e.g., in the QED
inclusive-production process $e^+e^- \to \mu^+ \mu^- +X$,
or even in the simpler process $\gamma \gamma \to \mu^+ \mu^- +X$). 
In summary, processes that at the leading order (LO)
are produced with an exactly vanishing value, $\qt = 0$, of the transverse
momentum $\qt$ of the system $F$ tend to develop sooner or later (in the computation of
QCD and QED radiative corrections at subsequent perturbative orders)
f.o. divergences.

A possible reaction to this state of affairs is to advocate non-perturbative
strong-interactions dynamics and related non-perturbative QCD effects that can
cure the f.o. divergences. This cure would spoil the common wisdom
according to which non-perturbative contributions to infrared and collinear safe
observables are (expected to be) power suppressed, namely, suppressed by some
inverse power of the relevant hard-scattering scale (as set by the high invariant
mass $M$ of the system $F$).
Moreover, non-perturbative strong-interactions dynamics cannot be advocated in the
case of f.o. divergences in a pure QED context.

We pursue a more conventional viewpoint within perturbation theory.
We examine the origin of the singularities for azimuthal-correlation observables
order by order in perturbation theory, and we proceed to resum the singular terms
to all orders \cite{Catani:2010pd, Catani:2014qha}.

The singularities originate from the kinematical region where $\qt \to 0$.
We identify two sources of singularities. One source 
\cite{Nadolsky:2007ba, Catani:2010pd} is initial-state {\em collinear} radiation 
in the
case of systems $F$ that can be produced with vanishing $\qt$ by
gluon 
initiated
partonic collisions. The other source \cite{Catani:2014qha} is {\em soft}
wide-angle radiation in the case of systems $F$ that contain colour-charged
particles (or electrically-charged particles for QED radiative corrections). 
Collinear radiation is, per se, 
responsible for singularities in $\cos(n\varphi)$ (and $\sin(n\varphi)$)
harmonics with $n=2$ and $n=4$, only \cite{Catani:2010pd}.
Soft radiation is responsible for singularities that, in principle, can affect
harmonics with any (both {\em even} and {\em odd}) values of $n$
($n=1,2,3,4\dots$).

At high perturbative orders collinear and soft radiation is mixed up, and the
singularities are enhanced by logarithmic ($\ln(M/\qt)$) contributions.
We discuss in general terms how the all-order resummation of the logarithmic
contributions leads to ($\qt$ integrated) azimuthal asymmetries that are finite
and computable. Such procedure is effective in both QCD and QED contexts.
In the QCD case this does not imply that non-perturbative effects have a
negligible quantitative role in the region of very-low values of $\qt$, but
these effects give small contributions to $\qt$-integrated azimuthal-correlation
observables.
Within perturbative QCD, the most advanced theoretical treatment of singular
azimuthal correlations that is available at present regards the process of
heavy-quark pair production \cite{Catani:2014qha}. We use the process of top-quark
pair production in $pp$ collisions at the LHC as a prototype to present a first
illustration of the resummation of singular azimuthal asymmetries at the
quantitative level.

A correspondence can be established 
\cite{Nadolsky:2007ba, Catani:2010pd, Echevarria:2015uaa}
between the TMD distribution of linearly-polarized gluons
and singular azimuthal correlations that are due to initial-state collinear
radiation in gluon initiated partonic subprocesses. Therefore, both the
framework used in Refs.~\cite{Pisano:2013cya, Boer:2017xpy} and our perturbative
QCD treatment lead to corresponding $\cos(2\varphi)$ and $\cos(4\varphi)$
azimuthal modulations. In this respect,
the TMD distribution of linearly-polarized gluons
can be regarded as the extension of specific features of QCD collinear dynamics
from perturbative to non-perturbative transverse-momentum scales.
Much discussion of the present paper is related to soft radiation effects that
occur in processes with colour-charged final-state particles. 
Soft radiation produces azimuthal asymmetries
(which are singular at f.o. and finite after resummation) 
for {\em both} quark
(or antiquark) and gluon initiated subprocesses, 
and for $\cos(n\varphi)$ harmonics with arbitrary even values 
($n=2,4,6,\dots$) \cite{Catani:2014qha}
and also odd values of $n$. These soft wide-angle radiation effects are
unrelated to the TMD distribution of linearly-polarized gluons.

The paper is organized as follows.
In Sect.~\ref{sec:corr} we start our discussion on azimuthal asymmetries and we
specify the conditions that lead to divergences in f.o. computations.
In Sect.~\ref{sec:exa} we discuss azimuthal asymmetries in two examples of hadron collider processes, i.e. the production of lepton pairs through the DY mechanism and the production of a $t{\bar t}$ pair, by contrasting the different
behaviour of the corresponding azimuthal harmonics.
In Sect.~\ref{sec:4} we start our analysis of the small-$q_T$ limit:
in Sect.~\ref{sec:qt} we focus on the small-$q_T$ behavior at f.o. and
in Sect.~\ref{sec:azav} we recall the transverse-momentum resummation procedure
for the case of azimuthally-averaged cross sections.
In Sect.~\ref{sec:azor} we discuss the origin of singular azimuthal correlations
and present illustrative lowest-order results for the $Z+$jet production
process. In Sect.~\ref{sec:resaz} we outline the resummation procedure of
singular terms in the
case of azimuthally-correlated cross sections and we contrast the small-$q_T$
behavior expected for the resummed cross section with the known behavior of the
azimuthally-averaged transverse-momentum cross section.
In Sect.~\ref{sec:ttbar} we focus on the $n=2$ harmonic for $t{\bar t}$ production and we present first quantitative results of a resummed calculation at 
next-to-leading logarithmic 
accuracy. After the matching with the complete 
next-to-leading order (NLO)
result, the resummed computation offers an effective `lowest-order' prediction for the 
$n=2$ harmonic. We also comment on the possible role of
non-singular
terms. In Sect.~\ref{sec:summa} we summarize our results.

\setcounter{footnote}{2}

\section{Azimuthal correlations and asymmetries in 
fixed-order perturbation theory}
\label{sec:corr}

Our discussion on azimuthal correlations has a high generality. 
To simplify the illustration of the key points, we consider  
the simplest class of processes, in which the produced high-mass
system
in the final state is formed by only two `particles' in generalized sense
(point-like particles and/or jets).

We consider the inclusive hard-scattering hadroproduction process
\begin{equation}
\label{fprod}
h_1(P_1) + h_2(P_2) \to F(\{ p_3, p_4 \}) + X \;\;,
\end{equation}
in which the collision of the two hadrons $h_1$ and $h_2$ with momenta
$P_1$ and $P_2$ produces the triggered final state $F$, and $X$ denotes the
accompanying final-state radiation.
The observed final state $F$ is a generic system that is formed by two 
`particles', $f_3$ and $f_4$, with four momenta $p_3$ and $p_4$, respectively.
The two particles can be point-like particles or hadronic jets ($j$), which are
reconstructed by a suitable (infrared and collinear safe) jet algorithm.
As for the case of point-like particles, the most topical process is the
production of a high-mass lepton pair $\ell \ell^\prime$ through the DY mechanism
of quark--antiquark annihilation. We consider many other cases such as, for
instance, the production of a photon pair ($\gamma \gamma$), a pair of top quark
and antiquark ($t {\bar t}$) or a pair of vector bosons ($VV$), in addition to
dijet production ($jj$) and associated production processes such as vector boson
plus jet ($Vj$). The invariant masses of the two particles have a little role in
the context of our discussion (and they do not affect any conceptual aspects
of our discussion). The system $F$ has {\em total} invariant mass $M$ 
($M^2= (p_3+p_4)^2$), transverse momentum $\bqt$ and rapidity $y$ (transverse
momenta and rapidities are defined in the centre--of--mass frame of the colliding
hadrons). We require that $M$ is large ($M \gg \Lambda_{QCD}$, $\Lambda_{QCD}$
being the QCD scale), so that the process in Eq.~(\ref{fprod}) can be treated
within the customary perturbative QCD framework. We use ${\sqrt s}$ to denote the 
centre--of--mass energy of the colliding hadrons, which are treated in the
massless approximation ($s=(P_1 + P_2)^2=2 P_1 \cdot P_2$).

The dynamics of the production process in Eq.~(\ref{fprod}) can be described in
terms of five kinematical variables: the total mass $M$, transverse momentum
$\qt$ and rapidity $y$ of the system $F$ and two independent angular variables
that specify the kinematics of the two particles $f_3$ and $f_4$ with respect to
the total momentum $q= p_3+p_4$ of $F$. These two angular variables are a
polar-angle variable and an azimuthal-angle variable. To be definite and to avoid
the use of `exotic' variables, we refer to a widely-used set of angular variables
and we use the polar angle $\theta$ and the azimuthal angle $\varphi$ (of particle
$f_3$) in the Collins--Soper (CS) rest frame\footnote{Since we are dealing with a
rest frame of $F$, the two particles are exactly (by definition) back--to--back in
that frame. In particular, the relative azimuthal separation is $\Delta \varphi =
\pi$.} \cite{Collins:1977iv} of the system $F$.

The variable $\varphi$ is the relevant variable for our discussion of 
azimuthal correlations. We remark that $\varphi$ specifies the azimuth of one of
the two particles in the system $F$ with respect to the 
total momentum of the system.
In particular, we also remark that we are not considering the relative azimuthal
separation $\Delta \phi= \phi_3 - \phi_4$ ($\phi_i=\phi(\bom{p_T}_i)$,
with $i=3,4$, is the azimuthal angle of the transverse-momentum vector
$\bom{p_T}_i$ in the centre--of--mass frame of the colliding hadrons) between the
two particles. However, we can anticipate (we postpone comments on this point)
that our main findings are not specific of the CS frame, and they are equally
valid for other azimuthal variables with respect to the system $F$
(for instance, we can consider the azimuthal angle in a different rest frame of
$F$ or, simply, the azimuthal difference $\phi(\bom{p_T}_i) - \phi(\bqt)$, where
$\phi(\bqt)$ is the azimuthal angle of $\bqt$ 
in the centre--of--mass frame of the colliding hadrons).

Using the kinematical variables in the CS frame, we can consider azimuthal
distributions for the process in Eq.~(\ref{fprod}). The most elementary
azimuthal-dependent observable is the azimuthal cross section at fixed invariant
mass,
\beq
\label{dphi}
\frac{d\sigma}{dM^2 \,d\varphi} \;\;,
\eeq
and we can also consider less inclusive observables such as, for instance,
the $\qt$-dependent azimuthal cross section in Eq.~(\ref{dphiq}) and the
multidifferential (five-fold) cross section in Eq.~(\ref{d5}):
\beq
\label{dphiq}
\frac{d\sigma}{dM^2 \,d\qt^2 \,d\varphi} \;\;,
\eeq
\beq
\label{d5}
\frac{d\sigma}{dM^2 \,dy \,d\qt^2 \,d\!\cos\theta \,d\varphi} \;\;.
\eeq
All these quantities are related (from the less inclusive to the more inclusive
case) through integration of kinematical variables (for instance, the azimuthal
cross section in Eq.~(\ref{dphi}) is obtained by integrating Eq.~(\ref{dphiq})
over $\qt^2$) and, in particular, the azimuthal integration of Eq.~(\ref{dphi})
gives the total cross section (at fixed invariant mass) of the process:
\beq
\label{tot}
\frac{d\sigma}{dM^2} = \int_0^{2\pi} d\varphi 
\;\,\frac{d\sigma}{dM^2 \,d\varphi} \;\;.
\eeq
Obviously, we can also consider differential cross sections that are integrated
over a certain range of values of the invariant mass.

Since all the cross sections that we have just mentioned are infrared and collinear
safe quantities \cite{Sterman:1977wj},
they can be computed perturbatively within the customary QCD factorization
framework (see Ref.~\cite{Collins:1989gx} and references therein).
The only non-perturbative (strictly speaking) input is the set of 
PDFs
of the colliding hadrons. The PDFs are convoluted
with corresponding partonic differential cross sections $d{\hat \sigma}$ that can
be evaluated as a power series expansion in the QCD running coupling $\as(M)$.

This perturbative QCD framework is applicable at any finite (and arbitrary) fixed
perturbative order. Despite this statement, the first main observation that we
want to make is that the f.o. perturbative calculation of the 
azimuthal distributions can lead to divergent (and, hence, unphysical and useless)
results. More specifically, the f.o. calculation of the azimuthal cross section of
Eq.~(\ref{dphi}) gives the following results:
\begin{equation}
\label{div}
\frac{d\sigma}{dM^2 \,d\varphi} =
\begin{cases}
\text{finite at any f.o. \;\;\;\;\;\quad\quad\quad\quad\quad\quad
(DY production)}\\
{}\\
\text{divergent for any $\varphi$ at some f.o. \;\;($t{\bar t},Vj,
jj, \gamma\gamma, ZZ, W^+W^-, ..,$ production) }
\end{cases}
\;.
\end{equation}
We mean that the perturbative computation of $d\sigma/dM^2 \,d\varphi$ for the DY
process gives a finite result order--by--order in QCD perturbation theory, while
in most of the other cases (some of them are listed in the right-hand side of
Eq.~(\ref{div})) the computation gives a divergent (meaningless) result for 
{\em any} values of the azimuthal angle $\varphi$ starting from {\em some}
perturbative order.
We note that the integration over $\varphi$ of $d\sigma/dM^2 \,d\varphi$ gives the
total cross section (see Eq.~(\ref{tot})), which is known to be finite at any 
f.o..
Therefore,  the divergent behaviour that is highlighted in Eq.~(\ref{div})
originates from the {\em azimuthal-correlation}\footnote{Using the shorthand
notation $d\sigma(\varphi)$ to denote a generic multidifferential cross section
with azimuthal dependence (e.g., the cross sections in Eqs.~(\ref{dphiq}) and 
(\ref{d5})), its azimuthal average is 
$\langle d\sigma(\varphi) \rangle_{\rm av.}$ and
we can define the corresponding correlation component as 
$d\sigma^{\rm corr}(\varphi) \equiv d\sigma(\varphi) - 
\langle d\sigma(\varphi) \rangle_{\rm av.}$, analogously to the definition in
Eq.~(\ref{corr}).} 
component,
$d\sigma^{\rm corr}/dM^2 \,d\varphi$, of the azimuthal cross section:
\beq
\label{corr}
\frac{d\sigma^{\rm corr}}{dM^2 \,d\varphi} \equiv 
\frac{d\sigma}{dM^2 \,d\varphi} - 
\langle \frac{d\sigma}{dM^2 \,d\varphi} \rangle_{\rm av.}
= \frac{d\sigma}{dM^2 \,d\varphi} - \frac{1}{2 \pi} \;\frac{d\sigma}{dM^2} \;\;,
\eeq
where the notation $\langle \cdots \rangle_{\rm av.}$ denotes the azimuthal
average and $d\sigma/dM^2$ is the total cross section in Eq.~(\ref{tot}).

To understand the origin of the divergent behaviour in Eq.~(\ref{div}), we first
comment about kinematics. If the system $F$ has vanishing transverse momentum
($\qt=0$), the rest frame of $F$ is obtained by simply applying a longitudinal
boost to the centre--of--mass frame of the colliding hadrons. Any additional
rotation in the transverse plane of the collision leaves the system $F$ at rest and
makes the particle azimuthal angle $\varphi$ ambiguously defined.
Owing to the azimuthal symmetry of the collision process, if $\qt=0$ there is no
preferred direction to define $\varphi$. In other words, considering the azimuthal
angle $\varphi$ (as defined in the CS frame or any other rest frame of $F$) and
performing the limit $\qt \to 0$, we have
\beq
\label{angles}
\cos \varphi = \cos (\phi_3 - \phi(\bqt)) + {\cal O}(\qt/M) \;\;,
\eeq
where $\phi_3=\phi(\bom{p_T}_3)$ and $\phi(\bqt)$ are the azimuthal angles of the
corresponding transverse-momentum vectors in the centre--of--mass frame 
of the colliding hadrons. If $\qt=0$, $\phi(\bqt)$ is not defined and,
consequently, $\varphi$ is not (unambiguously) defined.
At the strictly formal level, to define the azimuthal correlation we have to
exclude the phase space point at $\qt=0$. If we want to consider
$\qt$ integrated azimuthal correlations (such as, for instance, the cross section
in Eq.~(\ref{corr})), we have to introduce a minimum value of $\qt$
($\qt > q_{\rm cut}$) and eventually perform the limit 
$q_{\rm cut} \to 0$. The divergences that are highlighted in Eq.~(\ref{div})
do not appear by considering azimuthal correlations at fixed and finite values of 
$\qt$ (e.g., the azimuthal-dependent cross sections in Eq.~(\ref{dphiq}) and
(\ref{d5})). These  divergences are related to the limit $q_{\rm cut} \to 0$
after integration of the azimuthal correlations over $\qt$.

Our discussion on the limit $q_{\rm cut} \to 0$ reconciles the appearance of
divergences in Eq.~(\ref{div}) with the perturbative criterion of infrared and
collinear safety at the formal level: no divergence occurs provided 
$\qt \neq 0$, namely, provided the azimuthal-correlation observable is well 
(unambiguously) defined. However, a single phase space point at $\qt = 0$
is physically harmless. Therefore, the $\qt$ integrated azimuthal correlations
(i.e., their limiting behaviour as $q_{\rm cut} \to 0$) are finite and
measurable quantities. Moreover, they are also finite and measurable if 
$q_{\rm cut}$ is non-vanishing, or, equivalently if $\qt$ is not vanishing and
fixed at an arbitrarily small value.
The divergence in Eq.~(\ref{div}) implies that the $\qt$ dependent azimuthal
correlations become {\em singular} at small values of $\qt$, and that this
singular behaviour is {\em not integrable} over $\qt$ in the limit $\qt \to 0$.
This unphysical behaviour of f.o. perturbative QCD at small values of $\qt$ and
the divergence of the $\qt$ integrated azimuthal correlations certainly require
some deeper understanding to have a QCD theory of physically measurable
azimuthal correlations. As a possible shortcut, one can still use f.o.
perturbative QCD but avoid the region of small values of $\qt$. In this case one
still needs some understanding of the phenomenon in order to assess the extent
of the dangerous small-$\qt$ region where the f.o. predictions are `unphysical'
or, anyhow, not reliable at the quantitative level.

We note that the kinematical relation in Eq.~(\ref{angles}) implies that 
the azimuthal angle
in the CS frame has no privileged role in the context of our discussion of
azimuthal correlations in the small-$\qt$ region. The main features of the
small-$\qt$ behaviour of azimuthal correlations that are discussed in this paper
are basically unaffected by using azimuthal angles as defined in other rest
frames of the system $F$, or, by using other related definitions of azimuthal
angles. For instance, 
one can simply replace the CS frame angle $\varphi$ with one of the relative 
azimuthal
angles $\phi(\bom{p_T}_i) - \phi(\bqt)$ ($i=3,4$) in the centre--of--mass frame
of the colliding hadrons. Alternatively, one can use the 
difference
between the azimuthal angles 
(in the centre--of--mass frame of the colliding hadrons)
of the two transverse-momentum vectors $\bom{p_T}_3 - \bom{p_T}_4$ and $\bqt$.
All these definitions of the relevant (for our purposes)
azimuthal-angle variable turn out to be
equivalent (because of Eq.~(\ref{angles})) in the limit $\qt \to 0$
(or, at very small values of $q_{\rm cut}$).
In the following we continue to mainly refer ourselves to the azimuthal angle in
the CS frame, although all the basic features that we discuss are unchanged by
considering the other definitions of the azimuthal angle.

We also note that the discussion that we have presented so far about azimuthal
correlations can be generalized in a straightforward way to consider the case in
which the system $F$ is formed by more than two particles.
In the multiparticle case, we can simply examine azimuthal correlations that are
defined by using the azimuthal angles of the various particles in $F$ in a
specified rest frame of $F$ (such as the CS frame), by simply using the
various relative azimuthal angles  $\phi(\bom{p_T}_i) - \phi(\bqt)$
in the centre--of--mass frame of the colliding hadrons, or by using some other
related (i.e., equivalent in the limit $\qt \to 0$) azimuthal variables.
Since the multiparticle cases do not present any additional conceptual issues
with respect to the two-particle case, in the following 
(for the sake of technical simplicity)
we continue to mostly
consider
the case of systems $F$ that are formed by two
particles.

The f.o. perturbative divergences of the azimuthal correlations originate
from QCD radiative corrections due to inclusive emission in the
final state of partons with
low transverse momentum (soft and collinear partons). The origin is discussed in
more detail in the following Sections of the paper (see, in particular,
Sect.~\ref{sec:azor}). Since the azimuthal
correlations behave differently in different processes (as stated in 
Eq.~(\ref{div})), we anticipate here the conditions that produce the divergent
behaviour. 

Azimuthal correlations can have f.o. divergences if the final-state
system $F(\{p_3,p_4\})$ can be produced by the partonic subprocess
$c_1\, c_2 \to F$ (see also Eq.~(\ref{lopro}) and accompanying comments)
where
\beeq
\hspace*{-15mm}&& \bullet~~{\rm at~least~one~of~the~initial\!-\!state~colliding~partons} \;c_1 \;
{\rm and} \;c_2 \;{\rm is~a~gluon};
\label{a1}\\
\hspace*{-15mm}&& \bullet~~{\rm at~least~one~of~the~final\!-\!state~particles~with~momenta}
\,p_3 \;{\rm and} \,p_4 \;{\rm carries~colour~charge}.
\label{a2}
\eeeq

The conditions in (\ref{a1}) and (\ref{a2}) follow from a generalization of the
results in Refs.~\cite{Catani:2010pd, Catani:2014qha}.
The divergences arise from the computation of the QCD radiative correction to the
partonic process $c_1\, c_2 \to F$ (they do not arise in the computation of that
partonic subprocess itself!). Specifically, the f.o. divergences originate from
collinear-parton radiation \cite{Catani:2010pd}
in the case of the condition (\ref{a1})
and from 
soft-parton radiation \cite{Catani:2014qha}
in the case of the condition (\ref{a2}). We remark that one of the 
conditions in (\ref{a1}) and (\ref{a2}) is sufficient to produce the f.o.
divergences. In particular, the condition (\ref{a2}) {\em necessarily}
produces f.o. divergences, while the condition (\ref{a1}) produces divergences
with some `exception' (words of caution) in few specific cases (see below).
Having discussed the source of f.o. divergences in general terms, we can comment
on some specific processes.

The production of heavy-quark pairs such as, for instance, $t{\bar t}$ can occur
through the partonic subprocesses of quark-antiquark annihilation ($q{\bar q} \to
t{\bar t}$) and gluon fusion ($gg \to t{\bar t}$). One of these subprocesses has
initial-state gluons and, moreover, both $t$ and $\bar t$ have QCD colour charge.
Therefore, if $F=\{t{\bar t} \}$ both conditions (\ref{a1}) and (\ref{a2})
are fulfilled and the corresponding azimuthal correlations have f.o. divergences
(as stated in Eq.~(\ref{div})). The same reasoning and 
conclusions apply to the production of
$F=\{ V j\}$ and $F=\{j j\}$ (for instance, $F=\{ V j\}$ can be produced through
$qg \to V j$, where at the lowest-order $j=q$ carries QCD colour charge).
In the case of $F=\{ \gamma \gamma \}$ production the final-state photons do not
carry QCD colour charge. However, diphoton production at the
next-to-next-to-leading order (NNLO) can occur through the subprocess
$gg \to \gamma \gamma$ (the interaction is mediated by a quark loop): therefore,
due to the condition (\ref{a1}), the azimuthal correlations for diphoton
production diverge starting from the N$^3$LO computation.
The cases $F=\{ Z Z\}$
and $F=\{ W^+ W^- \}$ behave similarly to $F=\{ \gamma \gamma \}$, and they also
lead to azimuthal correlations that have f.o. divergences.

We consider the DY process, where $F=\{ \ell \ell' \}$ and the high-mass lepton
pair originates from the decay of a vector boson $V$ ($V \to \ell \ell'$).
The final-state leptons have no QCD colour charge and, therefore, the condition
(\ref{a2}) is not fulfilled. The partonic subprocesses $qg \to V(\ell \ell')$ and
${\bar q} g \to V(\ell \ell')$ are forbidden by colour conservation, and the
partonic subprocess $gg \to V(\ell \ell')$ is forbidden by the spin~1 of the
vector boson. The DY lepton pair can be produced through 
$q {\bar q} \to V(\ell \ell')$, but this partonic subprocess has no initial-state
gluons. Therefore, also the condition (\ref{a1}) is not fulfilled and the
azimuthal correlation for the DY process have no f.o. divergences in the
computation of QCD radiative corrections (as it is well known and recalled in
Eq.~(\ref{div})). However, we remark that f.o. divergences do {\em appear}
in the computation of QED (or, generally, electroweak) radiative corrections to
azimuthal correlations for the DY process (this important point is discussed in
more detail in Sect.~\ref{sec:exa}).

We can comment on the production of a system of colourless particles, such as 
$F=\{ \gamma \gamma \}$ or $F=\{ Z Z\}$, in the specific case (or, better,
within the approximation) in
which those particles originate from the decay of a spin-0 boson, such as the
Standard Model (SM) Higgs boson $H$ (e.g., $H \to \gamma \gamma$ or $H \to Z Z$).
In this case the production mechanism $gg \to H$
(followed by the $H$ decay) is allowed. However, due to the spin-0 nature of $H$,
the $H$ decay dynamically decouples from the $H$ production mechanism: the
angular distribution of the final-state particles ($\gamma \gamma$ or $Z Z$)
with momenta $p_3$ and $p_4$ is dynamically flat (it has no dynamical dependence
on the decay angles, since it can only depend on the Lorentz invariant
$2 p_3 \cdot p_4 = M^2$,  namely, on the invariant mass $M$ of the produced pair
of particles). As a consequence, although the condition (\ref{a1}) is fulfilled,
there are no azimuthal correlations in this specific case and, hence, there are
no accompanying f.o. divergences. The absence of azimuthal correlations follows
from the requirement that the two final-state particles are due to the decay of a
spin-0 boson. As a matter of principle at the conceptual level we note that,
considering the final-state system $F=\{ \gamma \gamma \}$ or $F=\{ Z Z\}$, the
various subprocesses that contribute to the production mechanism $gg \to F$
(subprocesses with and without an intermediate $H$, and corresponding
interferences) are not physically distinguishable (this is, strictly speaking,
correct 
although the applied kinematical cuts can sizeably affect the relative size 
of the various contributing subprocesses). As we have previously discussed,
the production of such systems without the intermediate decay of a spin-0 boson 
leads to azimuthal correlations with f.o. perturbative divergences. Therefore,
if the condition (\ref{a1}) is fulfilled, we can conclude that sooner or later
(in the computation of subsequent perturbative orders) QCD radiative corrections
produce non-vanishing azimuthal correlations and ensuing f.o. divergences.

To the purpose of studying azimuthal correlations and presenting corresponding
quantitative results, we find it convenient
to introduce harmonic components of azimuthal-dependent cross sections. In
particular, we define the $n$-th harmonic:
\beq
\label{nh}
\frac{d\sigma_n}{dM^2} \equiv \int_0^{2\pi} d\varphi \;\cos(n\varphi) \,
\int_0^{+\infty} d\qt^2 \;\frac{d\sigma}{dM^2 \,d\qt^2 \,d\varphi}
\;\Theta(\qt - q_{\rm cut}) \;\;, 
\;\;\;\;\;\;\quad q_{\rm cut}= r_{\rm cut} \,M \;\;,
\eeq
where $n$ is a positive integer $(n=1,2,3, \dots)$.
Note that, in view of our previous discussion on the origin of the divergences
that are mentioned in Eq.~(\ref{div}), we have introduced a minimum value
$q_{\rm cut}$ of $q_T$ ($q_T > q_{\rm cut}$). 
We usually consider values of $q_{\rm cut}$ that are
proportional to the invariant mass $M$ of the system $F$ 
($q_{\rm cut} = r_{\rm cut} M$, with a fixed parameter $r_{\rm cut}$), although
also fixed values of $q_{\rm cut}$ can be used. One can also consider $n$-th
harmonics of more differential cross sections (e.g., differential with respect to
$y$ and $\cos \theta$ as in Eq.~(\ref{d5})). The $n$-th harmonic
in Eq.~(\ref{nh}) is defined by using the weight function $\cos(n\varphi)$;
$n$-th harmonics with respect to the 
sine function can also be considered by 
simply replacing 
$\cos(n\varphi) \to \sin(n\varphi)$ in the integrand of Eq.~(\ref{nh}).
In particular, the knowledge of the harmonics with respect to both 
$\cos(n\varphi)$ and  $\sin(n\varphi)$ for {\em all} integer values of $n$
$(n=1,2,3, \dots)$ is equivalent to the complete knowledge of the 
azimuthal-correlation cross section in Eq.~(\ref{corr}).
Note that the $n$-th harmonic in Eq.~(\ref{nh}) is not a positive definite
quantity. Although the physical azimuthal cross section 
$d\sigma/dM^2 \,d\qt^2 \,d\varphi$ is positive definite, the weight function
$\cos(n\varphi)$ (or $\sin(n\varphi)$) has no definite sign.

The $n$-th harmonic with weight $\cos(n\varphi)$ (or $\sin(n\varphi)$)
gives a direct measurement of the $\cos(n\varphi)$ (or $\sin(n\varphi)$)
{\em asymmetry} of the azimuthal-dependent cross section.
The QCD computation of the $n$-th harmonic gives a finite result provided 
$q_{\rm cut}$ is not vanishing. On the basis of Eq.~(\ref{div}),
in the limit $q_{\rm cut} \to 0$ some harmonics (asymmetries) can be divergent if
computed at some f.o. in QCD perturbation theory. 
We cannot draw general conclusions on
harmonics with odd values of $n$ ($n=1,3,5,\dots$), but 
from the results in
Refs.~\cite{Catani:2010pd, Catani:2014qha} 
we know that harmonics with
{\em even} values of $n$ ($n=2,4,6,\dots$) can have f.o. divergences. 
Specifically, 
if the condition (\ref{a1}) is fulfilled the harmonics with $n=2$ and $n=4$
have f.o. divergences \cite{Catani:2010pd}, and 
if the condition (\ref{a2}) is fulfilled {\em all} the $n$-even 
($n=2,4,6,8$ and so forth) $\cos(n\varphi)$ asymmetries have f.o. divergences
\cite{Catani:2014qha}. In Sect.~\ref{sec:exa} we explicitly show quantitative
results on f.o. computations of harmonics for the DY and $t{\bar t}$ production
processes. In Sect.~\ref{sec:azor} we also show f.o. results for
$Vj$ production and we further comments on $n$-odd harmonics.

As mentioned in the Introduction and discussed in 
Sect.~\ref{sec:resaz}, the f.o. divergences of the azimuthal
asymmetries can be cured by a proper all-order perturbative resummation of
QCD radiative corrections. The resummed computation leads to azimuthal
correlations and asymmetries that are finite, as it is the case of the 
corresponding
physical (and measurable) quantities.

\section{Examples}
\label{sec:exa}

The lepton angular distribution of the DY process is a much studied subject both
experimentally and theoretically. 
Recent measurements of the lepton angular distribution for $Z$ production at LHC
energies (${\sqrt s}= 8$~TeV) are presented in 
Refs.~\cite{Khachatryan:2015paa,Aad:2016izn}, together with corresponding QCD
predictions from Monte Carlo event generators and f.o. calculations.
A very recent phenomenological study of the DY lepton angular distribution is
performed in Ref.~\cite{Lambertsen:2016wgj}.
As for previous literature on the subject, we mainly refer the reader to the list
of references in Refs.~\cite{Khachatryan:2015paa, Aad:2016izn,
Lambertsen:2016wgj}.

The QCD structure of the DY lepton angular distribution is well known, and it is
a consequence of the spin-1 nature of the vector bosons $(V=Z,\gamma^*,W^\pm)$
involved in the DY mechanism. To be definite we consider the DY process at the
Born level with respect to electroweak (EW) interactions. Within this Born level
framework, the DY differential cross section is expressed in terms of a leptonic
tensor (for the EW leptonic decay of the vector boson) and a hadronic tensor
(for the hadroproduction of the vector boson). The QCD production dynamics is
embodied in the hadronic tensor and the two tensors are coupled through the spin
(helicity) correlations of the vector boson. Owing to quantum interferences, we
are dealing with 9 helicity components of the five-fold differential cross
section in Eq.~(\ref{d5}) (the hadronic tensor is a $3\times 3$ helicity
polarization matrix, due to the 3 polarization states of the vector boson),
which, therefore, can be expressed \cite{Mirkes:1992hu} as a linear combination
of 9 spherical harmonics or, better, harmonic polynomials (the leptonic tensor
has a polynomial dependence of rank 2 on the lepton momenta) of the lepton angles
$\theta$ and $\varphi$ (we specifically use the CS frame).
For the illustrative purposes of our discussion of azimuthal correlations, it is
sufficient to consider the four-fold differential cross section 
$d\sigma/(dM^2\,dy\,dq_T^2\,d\varphi)$, which is obtained by integrating 
Eq.~(\ref{d5}) over $\cos \theta$. We write
\beq
\label{dyphi}
2\pi \,\frac{d\sigma_{DY}}{dM^2 \,dy \,d\qt^2 \,d\varphi} =
\frac{d\sigma_{DY}}{dM^2 \,dy \,d\qt^2} +
[d\sigma]_3 \,\cos \varphi +
[d\sigma]_2 \,\cos(2\varphi) +
[d\sigma]_7 \,\sin \varphi +
[d\sigma]_5 \,\sin(2\varphi) \;,
\eeq
where $d\sigma/(dM^2\,dy\,dq_T^2)$ is the DY differential cross section
integrated over $\varphi$, and the shorthand notation $[d\sigma]_I$ denotes
differential cross section components that only depend on $M, y, q_T$. The
azimuthal dependence of Eq.~(\ref{dyphi}) is {\em entirely} given by the
harmonics $\{ \cos\varphi, \cos(2\varphi), \sin\varphi, \sin(2\varphi) \}$
that are explicitly written in the right-hand side of Eq.~(\ref{dyphi}).
The subscript $I$ in $[d\sigma]_I$ is defined according to a customary notation
in the literature 
\cite{Chaichian:1981va, Mirkes:1992hu, Khachatryan:2015paa, Aad:2016izn},
such that $[d\sigma]_I \equiv C_I A_I(M,y,q_T) \times d\sigma/(dM^2\,dy\,dq_T^2)$,
where the functions $A_I(M,y,q_T)$ $(I=0,1,\dots,7)$ are known as `DY angular
coefficients' and $C_I$ are normalization factors (specifically, we have
$C_2=1/4, C_5=1/2, C_3=C_7=3\pi/16$). A point that we would like to remark is
that the QCD dependence of the azimuthal correlations of the DY cross section
involves {\em only} four harmonics. The five-fold cross section in
Eq.~(\ref{d5}) for the DY process is expressed in terms of 9 harmonic polynomials
of $\theta$ and $\varphi$: however, their dependence on $\varphi$ is given only
in terms of the four harmonics that also appear in Eq.~(\ref{dyphi}).

At LO in QCD perturbation theory, the DY cross section 
is due to the
partonic subprocess $q{\bar q} \to V (\to \ell \ell')$, which is of 
${\cal O}(\as^0)$. At this order in $\as$, there is no azimuthal dependence.
Azimuthal correlations start to appear at the
NLO\footnote{Throughout the paper we use the labels LO,
NLO, NNLO and so forth according to the perturbative order in which the
corresponding result (or partonic subprocess) contribute to the total cross section
of that specific process. In the case of azimuthal correlations, the N$^k$LO
result can effectively represent a QCD prediction at some lower perturbative 
order.} through the {\em tree-level} partonic processes at ${\cal O}(\as)$
($q{\bar q} \to V + g$ and its crossing related channels, such as 
$q g \to V + q$). At this order only the cross section components 
$[d\sigma]_3$ and $[d\sigma]_2$ in Eq.~(\ref{dyphi}) are not vanishing.
The azimuthal harmonics $\sin\varphi$ and $\sin(2\varphi)$ receive
non-vanishing contributions only starting from NNLO processes. 
More precisely, these 
non-vanishing contributions \cite{Hagiwara:1984hi}
are entirely due to {\em one-loop} absorptive (and time reversal odd) corrections
to the partonic subprocess $q{\bar q} \to V + g$ and its crossing related
channels. Azimuthal correlations up to NNLO were first computed in 
Ref.~\cite{Mirkes:1992hu}. Azimuthal-correlation results at N$^3$LO
can be obtained, in principle, by exploiting recent progress on the computation
of the NNLO corrections to
`$V+1$\,jet' 
production \cite{Boughezal:2015dva, Ridder:2015dxa}.

We recall that the cross section components $[d\sigma]_2$ and $[d\sigma]_3$
in Eq.~(\ref{dyphi}) are parity conserving, 
while $[d\sigma]_5$ and $[d\sigma]_7$
are parity violating. 
Moreover, $[d\sigma]_3$ vanishes in the limit of vanishing axial coupling of $V$
to either quarks or leptons. Therefore,
if $V=\gamma^*$, only the $\cos(2\varphi)$ harmonic
contributes in Eq.~(\ref{dyphi}), while in the case of the five-fold
differential cross section of Eq.~(\ref{d5}) there is an additional
non-vanishing azimuthal contribution that is proportional to 
$\sin(2\theta) \cos\varphi$ and it is due to $[d\sigma]_1$.

All the quantitative results that are presented in this paper refer to $pp$
collisions at the LHC energy $\sqrt s = 8$~TeV. In the QCD calculations we use
the set MSTW2008
\cite{Martin:2009iq}
of PDFs at NLO.

To present some illustrative results for the DY process we consider on-shell 
$Z$ production and its leptonic decay $Z \to e^+e^-$ in an electron--positron
pair. Our QCD calculation is performed by using the numerical Monte Carlo
program {\tt DYNNLO} \cite{Catani:2009sm}. 
The EW parameters
are specified in the $G_\mu$ scheme and we use the following values:
the Fermi constant is $G_F=1.1663787 \cdot 10^{-5}$~GeV$^{-2}$, the mass of the
$Z$ boson is $M_Z=91.1876$~GeV and the mass of the $W$ boson is 
$M_W=80.385$~GeV. We use equal values of the factorization ($\mu_F$) and
renormalization ($\mu_R$) scales and we set $\mu_F=\mu_R=M_Z$.

\begin{figure}[th]
\centering
\hspace*{-0.3cm}
\subfigure[]{
\includegraphics[width=3.3in]{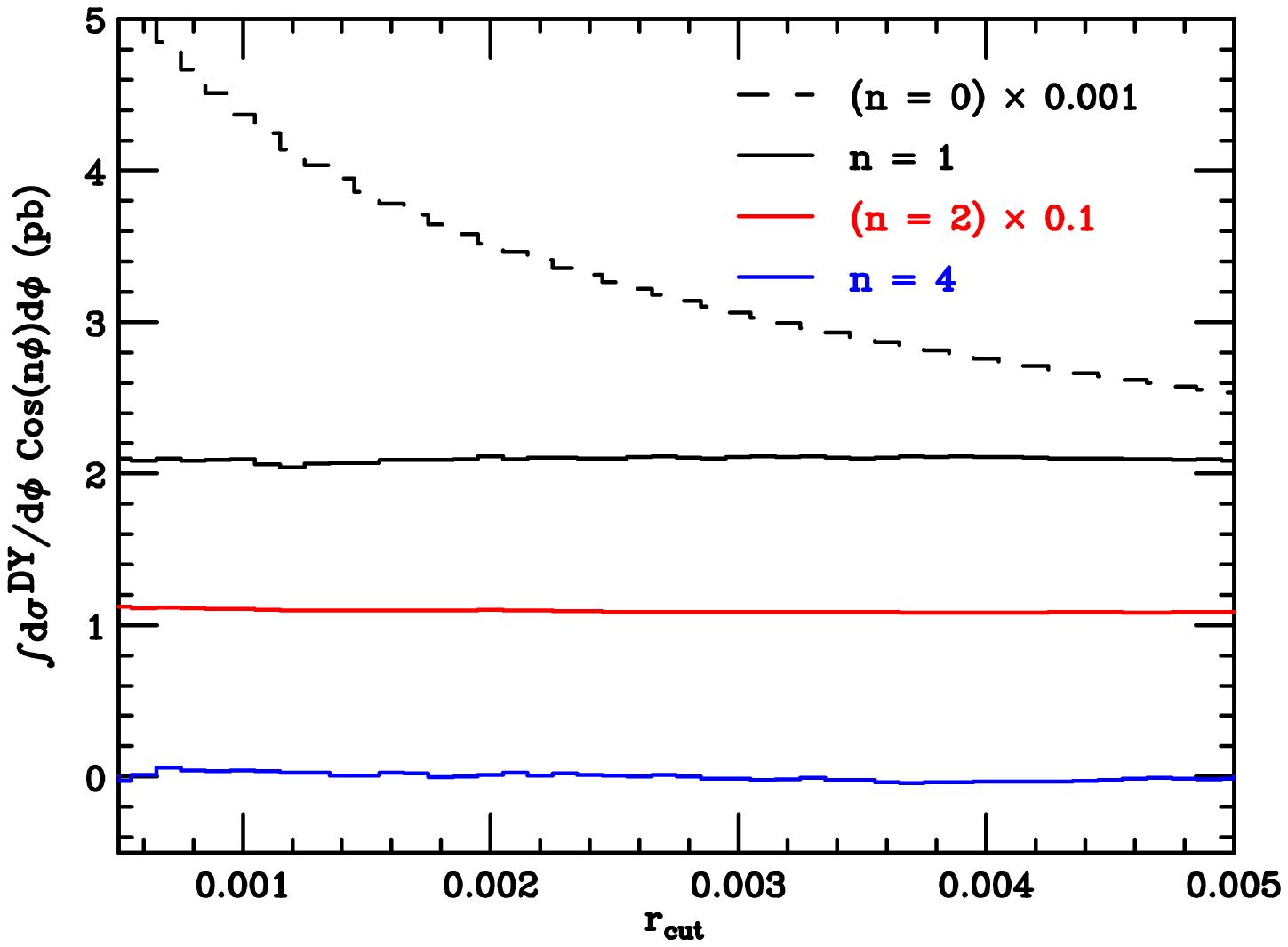}
}
\subfigure[]{
\includegraphics[width=3.4in]{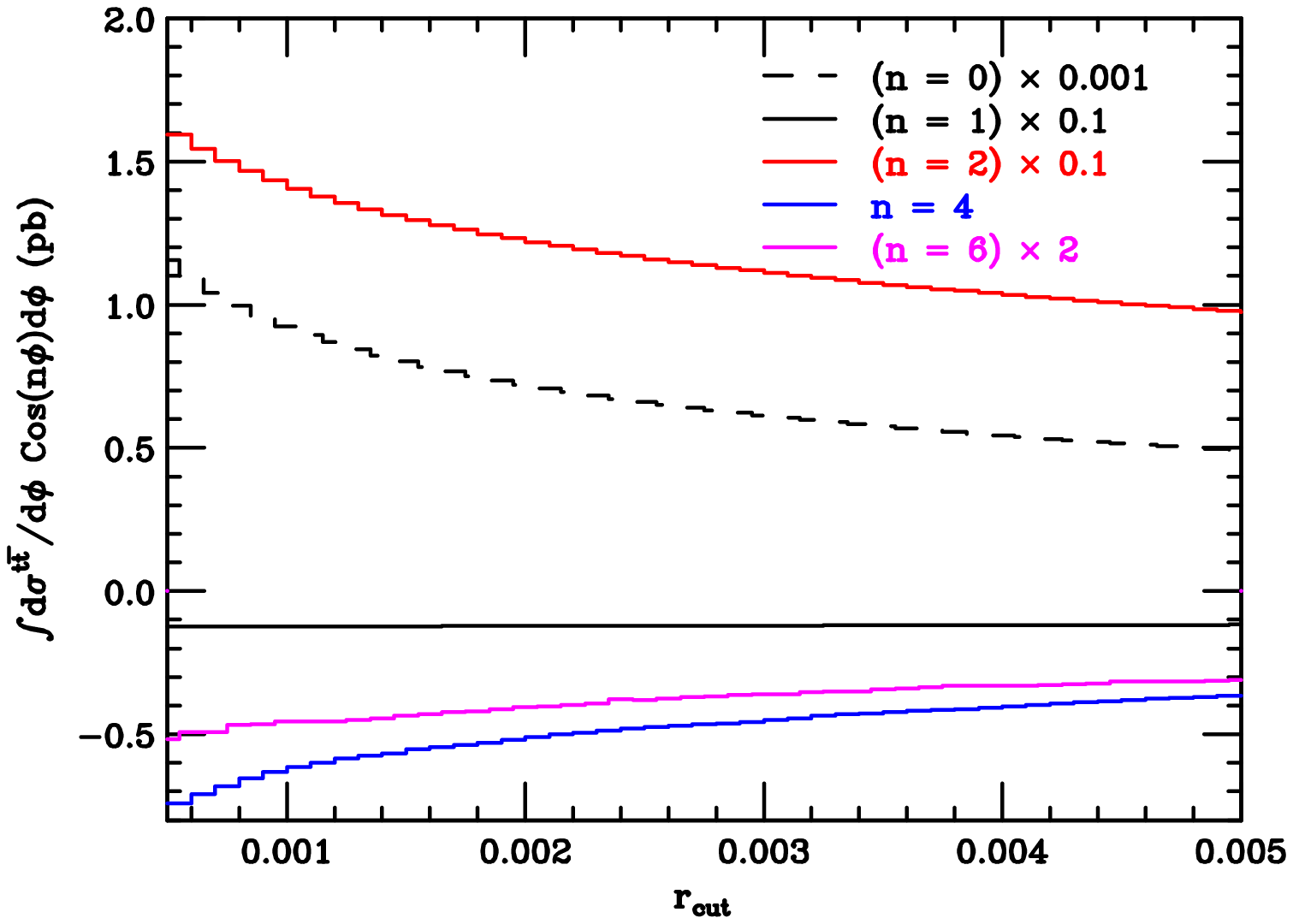}}
\caption{
\label{dyfig}
{\em Lowest-order harmonics as a function of $r_{\rm cut}=q_{\rm cut}/M$ in the cases of 
DY lepton pair (left) and $t\bar{t}$ (right) production at the LHC.}}
\end{figure}

We specifically consider the numerical calculation of various $\cos(n\varphi)$
asymmetries (see Eq.~(\ref{nh})), and $\varphi$ is the azimuthal angle of the
electron in the CS frame.
We evaluate the asymmetries at their non-trivial lowest order in f.o.
perturbation theory and, therefore, we compute all the NLO tree-level partonic
processes whose final state is `$Z(e^+e^-) + 1$~parton'. 
Our numerical results are presented in
Fig.~\ref{dyfig}-left.

The $n$-th harmonics are integrated over the lepton polar angle and
over the rapidity $y$ and transverse-momentum $q_T$ of the
$e^+e^-$ pair, and they are computed as a function of $r_{\rm cut}=q_{\rm
cut}/M_Z$, where $q_{\rm cut}$ is the minimum value of $q_T$ 
($q_T > q_{\rm cut}$). The results in Fig.~\ref{dyfig} are obtained for very
small values of $r_{\rm cut}$ in the range 
$5 \cdot 10^{-4} < r_{\rm cut} < 5 \cdot 10^{-3}$.
As we have already recalled, the azimuthal asymmetries for the DY process have
no f.o. divergences. Indeed, the results for $n=1,2,4$ that are presented in
Fig.~\ref{dyfig}-left show that the corresponding $\cos(n\varphi)$
asymmetries are basically independent of $r_{\rm cut}$ for 
very small values of $r_{\rm cut}$ and finite (the results in 
Fig.~\ref{dyfig}-left practically coincide with the numerical evaluation at 
$r_{\rm cut}=0$). The result for the $n=4$ asymmetry (blue line) in
Fig.~\ref{dyfig}-left is consistent with a vanishing value, in agreement with
the general expression in Eq.~(\ref{dyphi}) (the very small deviations that are
observed for $n=4$ in Fig.~\ref{dyfig}-left give an idea of 
the numerical uncertainties in
our calculation of the various asymmetries).
The result for the $n=1$ asymmetry (black line) gives a non-vanishing value (it
corresponds to the integral of the differential cross section component
$[d\sigma]_3$ in Eq.~(\ref{dyphi})). A non-vanishing value is obtained also for 
$n=2$ (red line), and it corresponds to the computation of the cross section
component $[d\sigma]_2$ in Eq.~(\ref{dyphi}). Note that the $n=2$ result 
reported in Fig.~\ref{dyfig}-left is rescaled by a factor of 0.1. Therefore, by
inspection of Fig.~\ref{dyfig}-left we can see that the $\cos(2\varphi)$
asymmetry is approximately a factor of 5 larger than the 
$\cos\varphi$ asymmetry.

In Fig.~\ref{dyfig}-left we also report the result for the $n=0$ harmonic
(dashed line) of `$Z + 1$~parton' production. The $n=0$ harmonic corresponds to the 
total (azimuthally-integrated) cross section, and it receives contributions from
both real and virtual emission subprocesses. Real and virtual terms are separately
divergent and their divergences cancel in the total contribution. In 
Fig.~\ref{dyfig}-left we report the result of the $n=0$ harmonic computed exactly
as specified in Eq.~(\ref{nh}), namely, by applying a non-vanishing lower limit 
$q_{\rm cut}= r_{\rm cut} M$ on $q_T$. Therefore, our computation selects only the
real-emission term due to the `$Z + 1$~parton' subprocesses.
As is well known (see also Sect.~\ref{sec:qt}), 
this real-emission
term diverges in the limit $q_{\rm cut} \to 0$ and the divergent behaviour is
proportional to $\ln^2 r_{\rm cut}$: in Fig.~\ref{dyfig}-left we clearly see the
increasing behaviour of the $n=0$ harmonic as $r_{\rm cut} \to 0$. In the actual
computation of the total cross section, the $n=0$ result in Fig.~\ref{dyfig}-left
has to be combined with the NLO real-emission term at still smaller values of 
$r_{\rm cut}$ and with the LO and NLO virtual terms, thus leading to a total
finite result.
For the sake of completeness, we report 
the value (including the numerical error of the Monte Carlo integration)
of the NLO total cross section $\sigma_{DY}^{NLO}$ that we obtain by using the same parameters as
used in the results of Fig.~\ref{dyfig}-left: it is
$\sigma_{DY}^{NLO}=1100 \pm 1$~pb.
We note that $\sigma_{DY}^{NLO}$ is roughly 100 times larger than the value of
the $\cos(2\varphi)$ asymmetry in Fig.~\ref{dyfig}-left.

The DY process is quite `special' with respect to azimuthal correlations: the
azimuthal correlations are finite order-by-order in QCD perturbation theory and
their general QCD structure involves only 4 azimuthal harmonics 
(as shown in Eq.~(\ref{dyphi})). For most of the other hard-scattering processes
(with few possible exceptions such as $H \to \gamma \gamma, ZZ$ production, as
discussed in Sect.~\ref{sec:corr}) azimuthal correlations behave differently:
usually they have an azimuthal dependence that involves an infinite set of
harmonics (all values of $n$) and in many cases f.o. QCD computations lead to
divergences.

We consider the production of heavy-quark $Q{\bar Q}$ pairs and we treat the heavy
quark and antiquark as on-shell particles. Our discussion equally applies to any
heavy quark, but we specifically consider top quarks since the on-shell treatment
is more suitable in this case. In Fig.~\ref{dyfig}-right we present results
for $t{\bar t}$ production that are obtained analogously to the DY results 
presented in Fig.~\ref{dyfig}-left.

Our QCD computation of $t{\bar t}$ production is performed by using the numerical
Monte Carlo program
of Ref.~\cite{Bonciani:2015sha}, which includes QCD radiative
corrections at NLO (it uses the NLO scattering amplitudes of the MCFM program
\cite{Campbell:2015qma}) 
and part of the NNLO contributions.
We set $\mu_F=\mu_R=m_t$ and we use the value $m_t= 173.3$~GeV for the mass $m_t$
of the top quark. We consider the azimuthal angle $\varphi$ of the top quark
in the CS frame, and in Fig.~\ref{dyfig}-right we present the numerical results
of various $\cos(n \varphi)$ asymmetries (see Eq.~(\ref{nh})) after their
integration over the invariant mass $M$ of the $t{\bar t}$ pair.
As in the case of the results in Fig.~\ref{dyfig}-left, the $t{\bar t}$ results
are integrated over the polar angle of the top quark and over the rapidity $y$
and transverse momentum $\qt$ ($\qt > q_{\rm cut}$) of 
the $t{\bar t}$ pair.
The azimuthal asymmetries are evaluated (as a function 
of $r_{\rm cut}= q_{\rm cut}/M$) at their non-trivial lowest order (i.e.,
${\cal O}(\as^3)$) in f.o. perturbation theory and, therefore, we compute all the
NLO {\em tree-level} partonic processes whose final state is 
`$t{\bar t} + 1$~parton'.

Before commenting on the results in Fig.~\ref{dyfig}-right, we recall
some of the results in Ref.~\cite{Catani:2014qha}. In Ref.~\cite{Catani:2014qha}
azimuthal correlations for $t{\bar t}$ production are computed in analytic form in
the small-$q_T$ limit. At ${\cal O}(\as^3)$ the result of 
Ref.~\cite{Catani:2014qha} shows that the $q_T$-dependent azimuthal-correlation
cross section $d\sigma^{\rm corr}/dq_T^2 d\varphi$ behaves proportionally to
$1/q_T^2$ in the limit $q_T \to 0$ and, hence, it is not integrable over 
$q_T$ down to the region where $q_T=0$. The $1/q_T^2$ behaviour is proportional to
a non-polynomial function of $\cos^2 \varphi$ that, therefore, leads to divergent
$\cos(n \varphi)$ harmonics for even values of $n$ $(n=2,4,6,\dots)$.
The $q_T$ spectra of harmonics with odd values of $n$ have instead a less singular
$q_T$ behaviour at ${\cal O}(\as^3)$
and they are integrable over $q_T$ in the limit $q_T \to 0$.

The numerical results in Fig.~\ref{dyfig}-right are consistent with the convergent
or divergent behaviour predicted in Ref.~\cite{Catani:2014qha}.
The harmonic with $n=1$ (black line) is not vanishing and basically independent of
$r_{\rm cut}$ for very small values of $r_{\rm cut}$. The sign of the
$n=1$ harmonic is negative (the $n=1$ harmonic of $\bar t$ would be positive,
analogously to the corresponding harmonic of the electron in the DY case of
Fig.~\ref{dyfig}-left), and its absolute size (note that it is rescaled by a factor
of 0.1 in Fig.~\ref{dyfig}-right) is roughly a factor of two smaller than the size
of the $n=1$ harmonic for $Z(e^+e^-)$ production (Fig.~\ref{dyfig}-left).
The harmonics with $n=2$ (red line), $n=4$ (blue line) and $n=6$ (magenta line)
in Fig.~\ref{dyfig}-right have instead an increasing (in absolute value) size for
small and decreasing values of $r_{\rm cut}$: this behaviour is consistent with
a $\ln r_{\rm cut}$ dependence, as expected from the analytical results in  
Ref.~\cite{Catani:2014qha}. The results for 
$n=2,4$ and 6 in Fig.~\ref{dyfig}-right have no straightforward quantitative
implications for physical azimuthal asymmetries since they refer to small
values of $r_{\rm cut}$ (and they eventually diverge in the limit 
$r_{\rm cut} \to 0$). Nonetheless we observe that the absolute magnitude of the
$n$-even $\cos(n \varphi)$ asymmetries decreases as $n$ increases.
As in the case of Fig.~\ref{dyfig}-left for the DY process, in 
Fig.~\ref{dyfig}-right we also present the ${\cal O}(\as^3)$ result of the $n=0$
harmonic (dashed line) for the real-emission process `$t{\bar t} + 
1$~parton'. Analogously to the DY process, the 
`$t{\bar t} + 1$~parton' contribution to the $n=0$ harmonic diverges in the limit
$r_{\rm cut} \to 0$, and its dominant behaviour at small $r_{\rm cut}$ is
proportional to $\ln^2 r_{\rm cut} + {\cal O}(\ln r_{\rm cut})$. At small values of  
$r_{\rm cut}$ the shape of the $r_{\rm cut}$-dependence of the $n=0$ result
is thus steeper than that of the results for the harmonics with $n=2,4$ and 6.
After combining real and virtual contributions, the $n=0$ harmonic gives the total
cross section.
The value (including the numerical error of the Monte Carlo integration)
of the NLO total cross section that we find is 
$\sigma_{t{\bar t}}^{NLO}=226.2 \pm 0.1$~pb.
We note that $\sigma_{t{\bar t}}^{NLO}$ is roughly 200 times larger than 
the absolute value of
the $\cos \varphi$ asymmetry ($n=1$) in Fig.~\ref{dyfig}-right.

As we have anticipated in Sect.~\ref{sec:corr} with a brief sentence, we expect
(and predict) the appearance of f.o. perturbative divergences in the computation
of QED radiative corrections to azimuthal correlations for the DY process.
To explicitly explain this point we consider some analogies of the DY and 
$t{\bar t}$ production processes. We have illustrated the ${\cal O}(\as^3)$
divergences of the $n$-even $\cos(n \varphi)$ asymmetries for 
$t{\bar t}$ production. Part of these divergences arise from the computation of
the partonic subprocess $q {\bar q} \to t{\bar t} +g$ and, specifically, they
arise from the kinematical region where the radiated final-state gluon $g$
is soft and
at wide angle with respect to the direction of the initial-state $q$ and 
${\bar q}$. From this kinematical region the $t{\bar t}$ spectrum
$d\sigma/dq_T^2 d\varphi$ receives an azimuthal-correlation contribution that
behaves as $1/q_T^2$ and that depends on the QCD colour charges of the final-state
$t$ and $\bar t$. Specifically, this contribution is proportional to the Fourier
transformation of the function ${\bf D}^{(1)}$ in Eq.~(36) of 
Ref.~\cite{Catani:2014qha}. The first-order QED radiative corrections to the DY
process involve analogous partonic processes, 
such as $q {\bar q} \to Z(e^+e^-) +\gamma$,
with a soft and wide-angle photon $\gamma$ that is radiated in the final state. These
photon radiative corrections produce f.o. QED divergences to azimuthal 
asymmetries for the DY process. The DY analogue of the $t{\bar t}$ function
${\bf D}^{(1)}$ in Eq.~(36) of Ref.~\cite{Catani:2014qha} is simply obtained by
replacing the QCD colour charges of $t$ and ${\bar t}$ with the QED electric
charges of the DY final-state leptons (by simple inspection of 
Eq.~(36) in Ref.~\cite{Catani:2014qha}, such replacement leads to a non-vanishing
result).

More generally, we remark that the expression in Eq.~(\ref{dyphi})
for the DY process is valid only at the Born level with respect to the EW
interactions (although the expression is valid at arbitrary orders in QCD
perturbation theory). Including QED radiative corrections, we conclude that the
azimuthal-correlation cross section $d\sigma_{DY}/dq_T^2 d\varphi$ for the DY
process receives contributions from $\cos(n \varphi)$ and $\sin(n \varphi)$
harmonics with arbitrary values of $n$
(not only $n=1,2$ as in Eq.~(\ref{dyphi})). Moreover, 
the lowest-order 
perturbative QED
computation of the $\cos(n \varphi)$ contributions with $n$-even 
already leads to a
singular (not integrable) behaviour in the limit $q_T \to 0$ and to ensuing
divergent azimuthal asymmetries upon integration over $q_T$. The divergences
are switched on by QED radiative corrections and can receive additional
contributions from
powers of $\as$ in the context of mixed QED$\times$QCD radiative corrections.
Obviously, similar divergences arise also in a pure QED context such as, for
instance, in the QED computation of the process 
$e^+e^- \to \mu^+ \mu^- +X$.
Eventually all these divergences originate from the QED analogue of the condition
(\ref{a2}): the QED divergences arise if `at least one the final-state particles
with momenta $p_3$ and $p_4$ carries non-vanishing electric 
charge'.

We note that lepton pairs with high invariant mass can also be produced
throughout the partonic subprocess $\gamma \gamma \to \ell^+ \ell^-$, in which the
initial-state colliding photons arise from the photon PDF of the colliding
hadrons.
Strictly speaking, lepton pairs that are produced in this way are not physically
distinguishable\footnote{The reasoning is somehow analogous to that in 
Sect.~\ref{sec:corr} about diboson systems that can be produced with or without
an intermediate Higgs boson. The high-mass lepton pair can be produced with or
without an intermediate vector boson.}
from those that are produced by the DY mechanism of quark--antiquark annihilation.
Since the ratio between the photon PDF and the quark (or antiquark) PDF is
formally of ${\cal O}(\alpha/\as)$ ($\alpha$ is the fine structure constant),
the subprocess $\gamma \gamma \to \ell^+ \ell^-$ can be regarded as a QED
radiative correction in the context of the DY process. We remark
that 
radiative corrections to $\gamma \gamma \to \ell^+ \ell^-$ also produce
divergences in the computation of the lepton azimuthal correlations.
These divergences (see Sect.~\ref{sec:azor}) originate from the abelian (photonic) analogue 
of the condition (\ref{a1}): the divergences arise if
`at least one of the initial-state colliding partons $c_1$ 
and $c_2$ is a photon'.

To cure the f.o. perturbative divergences of azimuthal correlations one may
advocate non-perturbative strong-interactions dynamics and related
non-perturbative QCD effects, which can be sizeable in the small-$q_T$ region.
However, in the case of $q_T$-integrated azimuthal correlations (see
Eq.~(\ref{corr})) the non-perturbative QCD dynamics should cancel divergent
terms proportional to some powers of $\as(M)$, and this would imply that 
non-perturbative QCD effects scale logarithmically with 
$M$ (i.e., these effects would not be suppressed by some power of
$\Lambda_{QCD}/M$ in the hard-scattering regime $M \gg \Lambda_{QCD}$), thus
spoiling not only the {\em finiteness} but also the {\em infrared safety}
of the azimuthal correlations. Moreover such non-perturbative cure of the
divergent behaviour cannot be effective in the case of QED radiative corrections
and, in particular, it cannot be conceptually at work in a pure QED context
since QED is well-behaved in the infrared (small-$q_T$) region.
In the following Sections we show that the problem of f.o. divergences in azimuthal
correlations has a satisfactory solution entirely within the context of
perturbation theory. Namely, the resummation of perturbative corrections to all
orders leads to ($q_T$ integrated) azimuthal asymmetries that are finite and
computable. Such solution is effective in both QCD and QED contexts, although in
the QCD case this does not imply that
the non-perturbative effects have a negligible
quantitative role in the small-$q_T$ region.

\section{The small-$\qt$ region}
\label{sec:4}
The f.o. divergences 
of azimuthal
correlations arise from the small-$\qt$ region and, eventually, from the $\qt$
behaviour at the phase space point where $\qt=0$.
From a physical viewpoint, however, a single (isolated) phase space point
is harmless and, in particular, non-vanishing azimuthal correlations
(which are defined with respect to the direction of $\qt$) cannot be measured if $\qt \sim 0$. These physical requirements imply that the $\qt$
dependence of azimuthal-correlation observables has to be sufficiently smooth
in the small-$\qt$ region and, in particular, in the limit $\qt \to 0$.

We specify more formally these smoothness requirements. Since physical
measurements cannot be performed at a single phase space point, 
we split the $\qt$ integration region in the two intervals
where $0 \leq \qt \leq q_{\rm cut}$ (the lower-$\qt$ bin) 
and $\qt \geq q_{\rm cut}$ (the higher-$\qt$ region).
We then consider a generic azimuthal-correlation observable
(e.g., multidifferential cross sections as in 
Eqs.~(\ref{dphiq}) and (\ref{d5}),
or $n$-th harmonics as in Eq.~(\ref{nh}))
and we denote by $\sigma_l(q_{\rm cut};\varphi)$ and  
$\sigma_h(q_{\rm cut};\varphi)$
the `cross sections' that are obtained by $\qt$ integration of the observable
over the lower-$\qt$ bin and the higher-$\qt$ region, respectively.
The smoothness requirements are
\beeq
\label{s1}
&&\lim_{q_{\rm cut} \to 0} \;\sigma_l(q_{\rm cut};\varphi) = 0 \;\;, \\
\label{s2}
&&\lim_{q_{\rm cut} \to 0} \;\sigma_h(q_{\rm cut};\varphi) =
\sigma_{\rm tot}(\varphi) \;\;,
\eeeq
where $\sigma_{\rm tot}(\varphi)$ denotes the `total' cross section
(i.e., the result obtained by integrating the observable over the entire 
region of $\qt$).
These requirements specify azimuthal correlations that are 
physically well behaved.
In particular, Eq.~(\ref{s1}) implies that non-vanishing azimuthal correlations
cannot be physically observed if $\qt \sim 0$, and 
Eq.~(\ref{s2}) implies that the total azimuthal correlation is 
physically observable. Note, however, that Eqs.~(\ref{s1}) and (\ref{s2})
do not specify how $d\sigma/d\qt^2 d\varphi$ exactly behaves in the limit $\qt
\to 0$. Nonetheless, Eqs.~(\ref{s1}) and (\ref{s2}) imply that the sole phase
space point at $\qt=0$ (where azimuthal-correlation angles are not defined)
has no relevant physical 
role (the lower-$\qt$ bin  has an analogous
harmless role if $q_{\rm cut}$ is sufficiently small).

The f.o. divergences of azimuthal-correlation observables are due to the fact
that $d\sigma/d\qt^2 d\varphi$ does not have a (`sufficiently') smooth
dependence on $\qt$ at small values of $\qt$ if this quantity is computed
order-by-order in perturbation theory. In particular, 
$\sigma_h(q_{\rm cut};\varphi)$ diverges in the limit $q_{\rm cut} \to 0$
and, consequently, $\sigma_{\rm tot}(\varphi)$ is not computable, whereas 
$\sigma_l(q_{\rm cut};\varphi)$ is definitely not computable (divergent) even if
$q_{\rm cut}$ has a finite value.

As we have observed in Sect.~\ref{sec:exa}, also the customary
azimuthally-integrated (or azimuthally-averaged) cross section (i.e., the
harmonic with $n=0$ in Eq.~(\ref{nh}) or Fig.~\ref{dyfig}) 
does not fulfil the requirements in 
Eqs.~(\ref{s1}) and (\ref{s2}) (both 
$\langle \sigma_l(q_{\rm cut};\varphi) \rangle_{\rm av.}$
and $\langle \sigma_h(q_{\rm cut};\varphi) \rangle_{\rm av.}$
can separately be divergent in the limit $q_{\rm cut} \to 0$).
Therefore, the f.o. calculation of azimuthally-integrated cross section 
can have difficulties (as is well known) in describing the detailed $\qt$ shape
in the small-$\qt$ region.
In contrast to azimuthal-correlations observables, however, 
both $\langle \sigma_l(q_{\rm cut};\varphi) \rangle_{\rm av.}$
and $\langle \sigma_h(q_{\rm cut};\varphi) \rangle_{\rm av.}$
are computable for finite values of $q_{\rm cut}$. In particular,
the total cross section 
$\langle \sigma_{\rm tot}(q_{\rm cut};\varphi) \rangle_{\rm av.}
= \langle \sigma_l(q_{\rm cut};\varphi) \rangle_{\rm av.}
+ \langle \sigma_h(q_{\rm cut};\varphi) \rangle_{\rm av.}$
is always finite and computable (for arbitrary non-vanishing values of 
$q_{\rm cut}$) within f.o. perturbation theory.

In the following sections we discuss the small-$\qt$ behaviour of cross
sections in f.o. perturbation theory and after all-order resummation.

\subsection{Perturbative expansion 
with azimuthal-correlation terms}
\label{sec:qt}

Among all the hadroproduction processes of the type in Eq.~(\ref{fprod}), 
we consider those that at the LO in perturbative QCD are produced
by the following partonic subprocesses:
\beq
\label{lopro}
c_1 + c_2 \to F(\{p_3,p_4\}) \;\;,
\eeq
where $c_1$ and $c_2$ $(c_i=q,{\bar q},g)$ are the initial-state massless
colliding partons of the two hadrons $h_1$ and $h_2$. In the case in which one or
both particles (with momenta $p_3,p_4$) in $F$ is a jet, the notation in
Eq.~(\ref{lopro}) means that the jet is replaced by a corresponding QCD parton.
Note that the LO process in Eq.~(\ref{lopro}) is an `elastic' production process,
in the sense that $F$ is not accompanied by any additional final-state 
radiation\footnote{As usual, the final-state collinear remnants of the colliding
hadrons are not denoted in the partonic process.}.
This specification is not trivial, since it excludes some processes from our
ensuing considerations.
Some processes are excluded because of quantum number conservation. For instance,
if $F$ includes two top quarks (not a $t{\bar t}$ pair) no LO process as in 
Eq.~(\ref{lopro}) is permitted by flavour conservation. Some other processes are
excluded because of their customary perturbative QCD treatment. For instance, if
$F$ contains a hadron, its QCD treatment requires the introduction of a
corresponding fragmentation function and, consequently, $F$ is necessarily
produced with accompanying fragmentation products in the final state. Therefore,
in the following we exclude the cases in which $F$ includes one or two 
hadrons\footnote{Specifically, our ensuing discussion does not apply to azimuthal
correlations of systems $F$ that contain hadrons with momenta $p_3$ or $p_4$
(whereas it applies to infrared and collinear safe jets, which physically
contain hadrons).}. All the processes that are explicitly listed in 
Eq.~(\ref{div}) (including the DY process) have LO partonic processes of 
the type in Eq.~(\ref{lopro}).

The small-$q_T$ behaviour of azimuthal correlations for the processes that we
have just 
specified
is {\em partly} related to the behaviour of $q_T$
differential cross sections that are integrated over the entire azimuthal-angle
region (or, equivalently, that are azimuthally averaged). The behaviour in the
azimuthally-integrated (averaged) case is well known (starting from some of the
`classical' studies for the DY process
\cite{Dokshitzer:hw, Parisi:1979se, Curci:1979bg, Kodaira:1981nh,
Collins:1984kg}),
and in our presentation we contrast it with the cases with divergent azimuthal
correlations. To remark the differences between the two cases
in general terms, we use a shorthand
notation, which can be applied to final-state systems $F$ with two or {\em more}
particles (a more refined kinematical notation can be found in 
Refs.~\cite{Catani:2010pd, Catani:2014qha}).
A generic multidifferential cross section (e.g., Eq.~(\ref{dphiq}), 
Eq.~(\ref{d5}) or related observables) with dependence on $q_T$ and on the 
azimuthal-correlation angle is simply denoted by $d\sigma/dM^2 d^2{\bqt}$,
where $\bqt$ is the transverse momentum of $F$. Additional kinematical variables
that are possibly not integrated (such as rapidities and polar angles) are not
explicitly denoted. The dependence on a generic 
(as discussed in Sect.~\ref{sec:corr}) azimuthal-correlation angle with respect
to $q_T$ is denoted through the dependence on the direction $\qh={\bqt}/q_T$
of the two-dimensional vector ${\bqt}$. In practice such dependence will occur
through functions of scalar quantities such as, for instance, 
$\qh \cdot {\bf p}_{3 {\bf T}}$. In particular, independent of $\qh$ means
absence of azimuthal correlations, and the azimuthally-integrated (averaged)
cross section corresponds to the azimuthal integration (average) with respect to
the direction $\qh$ of $\bqt$.

Owing to transverse-momentum conservation in the process of Eq.~(\ref{lopro}),
at the LO level $F$ is produced with vanishing $q_T$. Its corresponding LO
$\qt$ cross section is proportional to $\delta^{(2)}(\bqt)$,
\beq
\label{losigma}
\frac{d\sigma^{LO}}{dM^2 \,d^2{\bqt}} \propto \delta^{(2)}(\bqt) \;\;,
\eeq
and the proportionality factor is simply the LO total ($\qt$ integrated) cross
section.
In the presence of the LO sharp $\qt$ behaviour of Eq.~(\ref{losigma}), NLO QCD
radiative corrections are dynamically enhanced in the small-$\qt$ region.
The dynamical enhancement is due to 
QCD radiation of low
transverse-momentum partons ({\em soft gluons} or QCD partons that are {\em
collinear} to the initial-state colliding partons), and it has the following
{\em general} form:
\beq
\label{nlosigma}
\frac{d\sigma^{NLO}}{dM^2 d^2{\bqt}} \propto \delta^{(2)}(\bqt) + \as
\left\{ \left(
a_2 \left[ \frac{1}{\qt^2} \ln\left(\frac{M^2}{\qt^2}\right)\right]_+
+ a_1 \left[ \frac{1}{\qt^2} \right]_+ + a_0 \;\delta^{(2)}(\bqt)
+ \frac{a_{\rm corr}(\qh)}{\qt^2} \right)+ \dots
\right\} \;,
\eeq
where the dots on the right-hand side denote contributions that are less
enhanced (`non-singular') in the limit $\qt \to 0$.
The `coefficients' $a_i$ ($i=0,1,2$) and $a_{\rm corr}$ in Eq.~(\ref{nlosigma})
depend on the process and they are independent of $\qt$ (we mean they are 
independent of
the magnitude $\qt$ of the vector $\bqt$). The important point 
(see below) is that 
$a_{\rm corr}$ does depend on the direction $\qh$ of $\bqt$, whereas 
$a_i$ ($i=0,1,2$) do not depend on it. All these coefficients 
($a_1,a_2.a_3,a_{\rm corr}$) can depend on the other kinematical variables
(e.g., rapidities and polar angles).
The QCD running coupling $\as$ in Eq.~(\ref{nlosigma}) and in all the subsequent
formulae 
is evaluated at a scale of
the order of $M$ (e.g., we can simply assume $\as=\as(M)$).

The terms that we have explicitly written in the right-hand side of 
Eq.~(\ref{nlosigma}) scale as $1/\qt^2$ (modulo logs) in the limit $\qt \to 0$
(i.e., they scale as $1/\lambda^2$ under the replacement 
$\bqt \to \lambda \bqt$).
The non-singular terms (which are simply denoted by the dots) represent
subdominant (`power-correction') contributions in the limit $\qt \to 0$.
These are, for instance, terms of the type $M/\qt$ or, more generally, terms that
are relatively suppressed by some powers (modulo logs) of $\qt/M$.
Independently of their specific form and of their azimuthal dependence,
these non-singular terms have an {\em
integrable} and {\em smooth} behaviour in the limit $\qt \to 0$.
Because of these features, the non-singular terms  and the ensuing 
azimuthal-correlation effects are well
behaved in the small-$\qt$ region.
Note, however, that the non-singular terms produce azimuthal effects whose
actual size (and $\qt$ behaviour) depends on the specific definition of the azimuthal-correlation
angle (see Eq.~(\ref{angles}) and related comments in Sect.~\ref{sec:corr}).

The expression in Eq.~(\ref{nlosigma}) is the master formula that we use for our
discussion of the small-$\qt$ behaviour of azimuthal correlations at NLO and
higher orders. Considering the `singular' terms (the
NLO terms that scale as $1/\qt^2$, modulo logs, in the limit
$\qt \to 0$), the azimuthal dependence is entirely embodied in 
$a_{\rm corr}(\qh)$. More precisely, the azimuthal-correlation dependence has
been separated (as in Eq.~(\ref{corr})) and embodied in 
$a_{\rm corr}(\qh)$ that, therefore,
gives a vanishing result after azimuthal
integration. Using the notation in Eq.~(\ref{corr}) we have
\beq
\label{av0}
\langle a_{\rm corr}(\qh) \rangle_{\rm av.} = 0 \;\;.
\eeq
This azimuthal-correlation term is absent for the DY process
(i.e., $a_{\rm corr}(\qh)=0$ in this case), and in all the other processes
it has no effect by considering $\qt$ dependent but azimuthally-integrated
observables. The presence of such term produces a $q_T$ behaviour that
definitely differs from the behaviour studied in the `classical' literature
\cite{Dokshitzer:hw, Parisi:1979se, Curci:1979bg, Kodaira:1981nh,
Collins:1984kg}
on the small-$\qt$ region and on QCD transverse-momentum resummation.


The other (`classical') terms in Eq.~(\ref{nlosigma}) are proportional
to the coefficients $a_i$ ($i=0,1,2$), and the symbol $[~~]_+$ denotes the
(`singular')  `plus'-distribution, which is defined by its action onto any
smooth function $f(\bqt)$ of $\bqt$. The definition is
\beq
\label{plus}
\int_0^{+\infty} d^2\bqt \;\,f(\bqt) \;\left[ g(\bqt) \right]_+
\equiv
\int_0^{+\infty} d^2\bqt \,\left[ f(\bqt) - f({\bf 0}) \,\theta(\mu_0 - \qt) \right]
\;g(\bqt) \;\;,
\eeq
where $g(\bqt)=\frac{1}{\qt^2}\ln^m(M^2/\qt^2)$ ($m=0,1$) in
Eq.~(\ref{nlosigma}),
and $\mu_0$ is a scale of the order of $M$ (at the formal level, varying the value
of $\mu_0$ changes the plus-distributions and the coefficient $a_0$, so that the
right-hand side of Eq.~(\ref{nlosigma}) is unchanged).
The plus-distribution in Eq.~(\ref{plus}) is equivalently defined through a
limit procedure as follows:
\beq
\label{pluslim}
\left[ g(\bqt) \right]_+ = \lim_{q_0 \to 0}
\left[ \theta(\qt - q_0) \; g(\bqt) - \delta^{(2)}(\bqt) 
\int
d^2{\bf k_T} \; g({ \bf k_T}) \;\theta(\mu_0- k_T) \,\theta(k_T - q_0) \right]
\;\;.
\eeq
Note that the point at $\qt=0$ is at the border of the phase space, but, at the
strictly formal level, it is {\em inside} the phase space (at variance with the
case of azimuthal correlations, in which it is formally {\em outside} the phase
space). This is essential to make sense of the LO $\qt$ differential cross
section in Eq.~(\ref{losigma}) and of the corresponding LO total cross
section. This is also essential (see Eqs.~(\ref{plus}) and (\ref{pluslim}))
to transform the non-integrable $\qt$ behaviour of 
$\frac{1}{\qt^2}\ln^m(M^2/\qt^2)$ into an integrable plus-distribution over the
small-$\qt$ region.
The plus-distribution $[~g(\bqt)~]_+$ involves two terms: a term that is simply
proportional to the function $g(\bqt)$ and a contact term (the term proportional
to $f({\bf 0})$ in Eq.~(\ref{plus}) or, equivalently, the term proportional
to $\delta^{(2)}(\bqt)$ in Eq.~(\ref{pluslim})).
At the conceptual level, these two terms arise from a combination of real and
virtual radiative corrections, which are separately infrared divergent. Real (soft and
collinear) emission corrections to the LO process in Eq.~(\ref{lopro}) 
produce the non-integrable terms $\frac{1}{\qt^2}\ln^m(M^2/\qt^2)$, while the
corresponding virtual radiative corrections lead to the contact terms that
eventually regularize the divergence at $\qt=0$.

The azimuthal-correlation term in Eq.~(\ref{nlosigma}) is instead 
proportional to
the divergent (not integrable) function $1/\qt^2$, rather than to a
corresponding plus-distribution. 
At the formal level one may be tempted to transform the azimuthal-correlation
term into a plus-distribution through the replacement 
$a_{\rm corr}(\qh)/\qt^2 \to [~a_{\rm corr}(\qh)/\qt^2~]_+$,
but such a replacement would not be effective since
\beq
\label{noplus}
\left[ \frac{a_{\rm corr}(\qh)}{\qt^2} \right]_+ = 
\frac{a_{\rm corr}(\qh)}{\qt^2} \;\;,
\eeq
namely, the plus-prescription is not able to regularize the non-integrable
behaviour of azimuthal-correlation terms. The equality in Eq.~(\ref{noplus})
is a consequence of the fact that the contact term (see Eq.~(\ref{pluslim}))
in the corresponding plus-prescription identically vanishes:
indeed we have
\beq
\label{nocontact}
\delta^{(2)}(\bqt) 
\int
d^2{\bf {k_T}} \;  \frac{a_{\rm corr}(\kh)}{k_T^2}\;\theta(\mu_0- k_T) 
\,\theta(k_T - q_0) =0
\;\;,
\eeq
because the azimuthal integration of $a_{\rm corr}(\kh)$
gives a vanishing result\footnote{The integral in Eqs.~(\ref{pluslim}) and
(\ref{nocontact}) may be interpreted as the integral over the small-$k_T$
region of the loop momentum $\bf k_T$ of the virtual NLO correction. After
integration over the azimuthal angle of $\bf {k_T}$, only the azimuthal 
average gives a non-vanishing contribution
to Eq.~(\ref{pluslim}), while the
azimuthal-correlation effect in Eq.~(\ref{nocontact}) vanishes.}
(see Eq.~(\ref{av0})).
We note that a result analogous to Eq.~(\ref{noplus}), namely
\beq
\label{av0gen}
\left[ g_{\rm corr}(\bqt) \right]_+ = g_{\rm corr}(\bqt) \;\;,
\eeq
is valid for any azimuthal-correlation function $g_{\rm corr}(\bqt)$,
i.e., any function such that 
$\langle g_{\rm corr}(\bqt) \rangle_{\rm av.}~=~0$.
Therefore, in general we can conclude that azimuthal-correlation terms that are
not integrable in the limit $\qt \to 0$ cannot be regularized by the 
plus-prescription. This general conclusion and, in particular, the equality in
Eq.~(\ref{noplus}) have a simple conceptual interpretation in the context of
f.o. perturbation theory. Virtual radiative corrections to the process in
Eq.~(\ref{lopro}) have $\qt =0$ and, therefore, they cannot contribute to 
azimuthal-correlation terms. In particular, they cannot provide the
real-emission azimuthal-correlation term with a (non-vanishing) contact term
that transforms the non-integrable real-emission contribution into an integrable
plus-distribution. In summary, the azimuthal correlation of the $\qt$ cross
section is a phenomenon that is necessarily produced by real emission (since
azimuthal dependence requires $\qt \neq 0$) and the f.o. divergences of the
azimuthal asymmetries are the consequence of a {\em complete} mismatch with virtual
emission, which is completely absent (i.e., it does not contribute) at the
{\em corresponding} perturbative order.
In the case of azimuthal-independent (plus-distribution) terms, instead, both real
and virtual effects contribute at the NLO, although their relative contribution is
highly unbalanced in the small-$\qt$ region.

The $r_{\rm cut}$ ($q_{\rm cut}$) dependence of the results that are presented
in Fig.~\ref{dyfig} is fully consistent with the small-$\qt$ behaviour in
Eq.~(\ref{nlosigma}). The harmonics with $n \neq 0$ receive contributions only
from the term that is proportional to $a_{\rm corr}(\qh)$ in 
Eq.~(\ref{nlosigma}), while $a_{\rm corr}(\qh)$ does not contribute to the
$n=0$ harmonics (because of Eq.~(\ref{av0})). We simply note that in the DY
process we have $a_{\rm corr}(\qh)=0$, whereas in the case of $t{\bar t}$
production $a_{\rm corr}(\qh)$ is not vanishing (in particular, the $n=1$
harmonic of $a_{\rm corr}(\qh)$ vanishes, while  $a_{\rm corr}(\qh)$ has
non-vanishing harmonics with $n=2,4,6$).
We also note that the results for the harmonics with $n=0$ (in both the DY and
$t{\bar t}$ production processes) follow from the $\qt$ shape of the
plus-distribution terms in Eq.~(\ref{nlosigma}): the $\qt$ integration of the
plus-distributions over the region $\qt > q_{\rm cut}= r_{\rm cut} M$ produces
the $r_{\rm cut}$ dependence in Fig.~\ref{dyfig}, and this 
$r_{\rm cut}$ dependence is exactly cancelled by that produced
from the integration of the 
plus-distributions over the region $0 \leq \qt \leq q_{\rm cut}$.

\setcounter{footnote}{2}

We include a comment on an additional point related to the f.o. perturbative
expansion of generic hard-scattering processes. In our discussion of
Eq.~(\ref{nlosigma}), we have considered processes that at the LO are initiated
by the partonic subprocesses in Eq.~(\ref{lopro}). We note that elastic
production processes of the type in  Eq.~(\ref{lopro}) can also appear at some
higher perturbative orders. For instance, as already mentioned in
Sect.~\ref{sec:corr}, this is the case if $F=\{ \gamma \gamma \},
\{ Z Z \},\{ W^+ W^- \}$. These diboson production processes are initiated at LO
by $q {\bar q} \to F$, and they also receive contributions from $ gg \to F$
(the gluons are coupled to $\gamma \gamma, Z Z, W^+ W^-$ through a quark
loop\footnote{The partonic process in Eq.~(\ref{lopro}) does not need to be a
tree-level process.}) starting from the NNLO. In these cases, the next-order
radiative corrections to $ gg \to F$ produce a small-$\qt$ behaviour that has
exactly the same structure as that in the round-bracket term of 
Eq.~(\ref{nlosigma}). Our following discussion throughout the paper equally
applies to all the processes that can occur through subprocesses of the type in
Eq.~(\ref{lopro}) at some perturbative order (different types of colliding
partons $c_1$ and $c_2$ can be involved at different perturbative orders).

Small-$\qt$ singular terms of the type in Eq.~(\ref{nlosigma}) are present in
the computation of $d\sigma/dM^2 d^2{\bqt}$ at each higher perturbative orders.
The $1/\qt^2$ behaviour of both azimuthally-independent and
azimuthally-correlated terms is enhanced by logarithmic factors,
$\ln^k(M^2/\qt^2)$, produced by multiple
radiation of soft and collinear partons.
There are at most two additional powers of $\ln(M^2/\qt^2)$ for each additional
power of $\as$, so that the dominant singular terms have a double-logarithmic
structure. The N$^k$LO radiative corrections to $d\sigma/dM^2 d^2{\bqt}$
include singular terms that are proportional to 
$\as^k \frac{1}{\qt^2}\ln^m(M^2/\qt^2)$ (with $m \leq 2k-1$), and this $\qt$
dependence is regularized by a plus-prescription {\em only} in the case of
azimuthally-independent contributions. Owing to Eq.~(\ref{av0gen}), the
plus-prescription is not effective for azimuthal-correlation terms.

The order-by-order singular behaviour that we have just discussed poses no
problems for a basic quantity such as the total ($\qt$ and azimuthally
integrated) cross section: the singular azimuthal-correlation terms cancel
because of the azimuthal integration, and the plus-distribution terms are
integrable over $\qt$.
As is well known, the plus-distribution terms (which remain after azimuthal
integration) are not `harmless' for differential cross sections. They lead to a
sharp unphysical behaviour of the small-$\qt$ differential cross section at the
NLO, and to large radiative-correction effects at any subsequent perturbative
order, with an unstable f.o. expansion and poor predictivity for the detailed
shape of the $\qt$ cross section in the small-$\qt$ region.
The smooth physical behaviour of the $\qt$ cross section and the predictivity of 
perturbative QCD can be recovered by the all-order resummation
(transverse-momentum resummation) of the logarithmically-enhanced
plus-distribution terms.

In the case of azimuthally-sensitive observables, the disease of f.o.
perturbation theory is more serious. The singular azimuthal-correlation terms
are not integrable and, as discussed in Sect.~\ref{sec:corr},  basic quantities
such as the total ($\qt$ integrated) cross section at fixed azimuthal angle 
(see Eq.~(\ref{div})) and the total ($\qt$ integrated) azimuthal asymmetries
(see Eq.~(\ref{nh})) can be divergent if they are evaluated in f.o. perturbation
theory. In the presence of divergences, also the f.o. perturbative result for 
$d\sigma^{\rm corr}/dM^2 d^2{\bqt}$ at fixed (and finite) values of
$\qt$ has an unphysical shape in
the small-$\qt$ region, since the f.o. result follows a non-integrable
(unphysical) behaviour.

In the following we briefly recall known results on transverse-momentum
resummation for `azimuthally-insensitive' observables. More precisely, we simply
sketch some main points of transverse-momentum resummation that are relevant for
our subsequent discussion on
azimuthal correlations and asymmetries.

\subsection{Transverse-momentum resummation with no divergent azimuthal
correlations}
\label{sec:azav}

In this subsection we limit ourselves to considering only singular contributions
to $d\sigma/dM^2 d^2{\bqt}$ that are independent of the azimuthal-correlation
angles. Order-by-order in perturbation theory, these contributions can be
{\em additively} separated from non-singular terms 
(see, e.g., the `dots' in the
right-hand side of Eq.~(\ref{nlosigma})), and they can also be 
{\em additively} separated from singular azimuthal-correlation
terms (see Eqs.~(\ref{corr}), (\ref{nlosigma}), (\ref{av0gen}) and accompanying
comments).
Therefore, we are dealing with {\em all} the singular plus-distribution terms
that appear in Eq.~(\ref{nlosigma}) and in corresponding higher-order
contributions.
As a consequence of the additive separability, azimuthally-insensitive,
azimuthally-integrated and azimuthally-averaged contributions have the same
equivalent meaning in the context of the discussion in this subsection.
In the case of processes whose azimuthal correlations are not singular
(such as the DY process in the context of QCD radiative corrections), 
the singular terms that we are
considering are the entire singular contributions to those processes.

Considering hard-scattering observables that are insensitive to divergent
azimuthal correlations, the all-order QCD treatment of the singular
contributions to the $\qt$ differential cross section in the small-$\qt$ region
is conceptually well known  
\cite{Parisi:1979se, Curci:1979bg, Collins:1984kg}).
In the cases in which the produced high-mass system $F$ is formed by particles
that carry no QCD colour charge (colourless system $F$), for instance, in the
cases of the DY process and of SM Higgs boson 
production, 
the treatment
is fully developed at arbitrary logarithmic accuracy by using various methods
and formalisms, such as direct QCD resummation
(see Ref.~\cite{Catani:2013tia}
and corresponding references therein),
Soft Collinear Effective Theory (SCET) methods (see, e.g.,
Refs.~\cite{Mantry:2009qz, Becher:2010tm}),
and transverse-momentum dependent (TMD) factorization (see, e.g.,
Refs.~\cite{Collins:2011zzd, GarciaEchevarria:2011rb, Collins:2012uy}).
In the case of processes whose final-state system $F$ includes colour-charged
particles (colourful system $F$), 
the theoretical treatment is formally less advanced 
and only few specific cases have been treated beyond the leading-logarithmic
(LL) level. For the specific process of heavy-quark pair production
($F=\{Q {\bar Q}\})$, transverse-momentum resummation has been explicitly
developed 
\cite{Zhu:2012ts, Catani:2014qha}
up to next-to-next-to-leading logarithmic (NNLL) accuracy.
The process of dijet production ($F=\{j j\}$) has been explicitly studied 
\cite{Sun:2014gfa}
up to next-to-leading logarithmic (NLL) accuracy
within the approximation of
small values of the cone size $R$ of the jets.

To the purposes of our subsequent discussion on azimuthal correlations,
we limit ourselves to recalling some features of the transverse-momentum
resummation formalism.
Transverse-momentum resummation has to be carried out \cite{Parisi:1979se}
in impact parameter
$(\bf b)$ space to exactly implement the relevant constraint
of transverse-momentum conservation for all-order multiparton radiation in the
inclusive final state.
The terms with plus-distributions (as in Eq.~(\ref{nlosigma})) at small values of
$\qt$ become powers of logarithms, $\ln (bM)$, in $\bf b$ space at large values
of $b$ ($bM \gg 1$). These logarithmically-enhanced contributions can be resummed
in $\bf b$ space.
Then the $\qt$ cross section
$d\sigma/dM^2 d^2{\bqt}$ is obtained by inverse Fourier transformation
(from $\bf b$ space to $\bqt$ space) of the all-order resummed result
in $\bf b$ space. Since we are dealing with azimuthally-insensitive
contributions to the $\qt$ cross section, the $\bf b$ space resummed expression
does not depend on the azimuthal angle of $\bf b$, and the inverse 
Fourier transformation can be recast in the form of a Bessel transformation.
The final result of the resummation procedure has the following (sketchy) form:
\beq
\label{resav}
\frac{d\sigma^{(\rm res)}_{\rm az. in.}}{dM^2 d^2{\bqt}} \;\propto \;
\frac{d\sigma^{(\rm res)}_{\rm az. av.}}{dM^2 d\qt^2} =
\int_0^{+\infty} db \; b \;J_0(b\qt) \;\Sigma_{\rm az. av.}^{(\rm res)}(M,b)
\;\;,
\eeq
where $J_0(x)$ is the 0th-order Bessel function, and the superscript `res'
denotes the resummed contribution to the $\qt$ differential cross section.
Since we are dealing with the resummation of the azimuthally-insensitive
(independent) plus-distribution terms, we have introduced the subscript 
`az. in.', and we simply point out that these terms equally contribute to the
azimuthally-averaged (`az. av.') $\qt$ cross section.

The integrand $\Sigma_{\rm az. av.}^{(\rm res)}(M,b)$ in Eq.~(\ref{resav}),
which is proportional to the LO total ($\qt$ integrated) cross section, embodies
the effect of the all-order perturbative resummation procedure in $b$ space.
The higher-order terms that are included in $\Sigma_{\rm az. av.}^{(\rm res)}$
are proportional to 
$\as^k \ln^{m+1}(b^2M^2)$, with $m \leq 2k-1$.
The dependence of $\Sigma_{\rm az. av.}^{(\rm res)}$ on the other kinematical
variables (e.g., rapidities and polar angles)
is not explicitly denoted.

The $b$ space cross section $\Sigma_{\rm az. av.}^{(\rm res)}$ involves
process-dependent and process-independent factors
(see, e.g., Refs.~\cite{Collins:1984kg, Catani:2010pd, Catani:2014qha}).
One of them is the Sudakov form factor $S_c(M,b)$
$(c=q,g)$ of the colliding partons $c_1$ and $c_2$ of the partonic subprocess in
Eq.~(\ref{lopro}).
The Sudakov form factor is universal (it is independent of $F$) and it includes
the entire effect of the resummation of the dominant double-logarithmic (DL)
contributions (from radiation of partons that are both soft and collinear to the
initial-state colliding partons) in $b$ space.

Within the DL approximation, the Sudakov form factor has the exponential
form\footnote{To be precise the DL behaviour of the Sudakov form factor
is $S(M,b)= \exp \{-2a_2 \as \ln^2(bM) \}$, where $a_2$ is the NLO
coefficient in Eq.~(\ref{nlosigma}).}
$e^{-\as \ln^2(bM)}$ \cite{Parisi:1979se} that produces a very strong damping
of the large-$b$ region ($bM \gg 1$) in the integrand on the right-hand
side of Eq.~(\ref{resav}). The suppression of the large-$b$ region
is stronger than that produced by any power of $1/(bM)$. The damping effect of
the Sudakov form factor eventually leads to resummed perturbative predictions for
the $\qt$ cross section that are physically well behaved (with a smooth $\qt$
dependence) in the small-$\qt$ region: the shape of the $\qt$ cross section
is the result of a (computable) smearing of the LO 
$\delta$-function
behaviour 
in Eq.~(\ref{losigma}).
In particular, the qualitative behaviour of 
$d\sigma^{(\rm res)}_{\rm az. av.}/dM^2 d\qt^2$
at very low values of $\qt$ can be examined by performing the limit $\qt \to 0$
of Eq.~(\ref{resav}). Using\footnote{The small-$\qt$ approximation 
$J_0(b\qt)= 1 +{\cal O}(b\qt) \simeq 1$ cannot be used if 
$\Sigma_{\rm az. av.}^{(\rm res)}(M,b)$ is perturbatively expanded
order-by-order. In this case there is no damping of the large-$b$ region and the
terms of ${\cal O}((b\qt)^k)$ in the expansion of the Bessel function cannot be
neglected (actually, it is the oscillatory behaviour of $J_0(b\qt)$
at $b\qt \gg 1$ that makes the f.o. expansion of 
$\Sigma_{\rm az. av.}^{(\rm res)}$ integrable over the region of large values of $b$).} 
$J_0(b\qt)= 1 +{\cal O}(b\qt)$ 
in Eq.~(\ref{resav}), we simply have 
\cite{Parisi:1979se}
\beq
\label{avqt0}
\frac{d\sigma^{(\rm res)}_{\rm az. av.}}{dM^2 d\qt^2} \propto {\rm const.} \;\; 
\quad \quad \quad (\qt \to 0) \;\;.
\eeq
Obviously our discussion at the qualitative level is much simplified
(though correct) for quantitative purposes. The quantitative size of the
resummation effects depends on the (classes of) subdominant logarithmic
contributions and, at very low values of $\qt$, it can also depend on
non-perturbative\footnote{Non-perturbative effects typically enhance the Sudakov
form factor suppression of the large-$b$ region, so that the qualitative
behaviour in Eq.~(\ref{avqt0}) is left unchanged.} effects (see, e.g.,
Ref.~\cite{Collins:va}) 
that affect the region of very large values of $b$ ($b \gtap 1/\Lambda_{QCD}$).

We note that experimental results on $\qt$ cross sections are usually presented
in terms of the differential cross section $d\sigma/d\qt$, rather than 
$d\sigma/d\qt^2$. The two cross sections are directly related by just a factor of
$2\qt$ from kinematics ($d\sigma/d\qt = 2\qt d\sigma/d\qt^2$). 
The azimuthally-averaged (or azimuthally-integrated) cross section
$d\sigma_{\rm az. av.}/d\qt$ has a peak in the small-$\qt$ region and 
it vanishes linearly in
$\qt$ as $\qt \to 0$: this behaviour is the consequence of the combined effect 
of
the dynamical small-$\qt$ behavior in Eq.~(\ref{avqt0}) and of the kinematical
suppression factor $2\qt$ as $\qt \to 0$.

We have previously noticed (in Sects.~\ref{sec:corr} and \ref{sec:qt})
that in many processes the final-state system $F$ can be elastically produced
(see Eq.~(\ref{lopro})) by subprocesses with different initial-state partonic
channels.
Our discussion about resummation equally applies to all these processes: the
corresponding $\qt$ cross section has several components (one component for each
contributing partonic channel) 
(see, e.g., Refs.~\cite{Catani:2010pd, Catani:2014qha}) and each component has
basically the same structure as in Eq.~(\ref{resav}).

\section{Azimuthal asymmetries and resummation}

\subsection{Origin of divergent azimuthal correlations}
\label{sec:azor}

In the context of QCD perturbation theory, singular azimuthal correlations in
the small-$\qt$ region were observed in the theoretical studies of
Refs.~\cite{Nadolsky:2007ba, Catani:2010pd, Catani:2014qha}.

Reference \cite{Nadolsky:2007ba} deals with diphoton production 
($F=\{ \gamma \gamma\}$) 
in hadron--hadron collisions. Considering the next-order radiative
corrections (which are part of the complete N$^3$LO corrections for diphoton
production) to the gluon fusion subprocess $gg \to \gamma \gamma$, the authors
of Ref.~\cite{Nadolsky:2007ba} find that the $\cos(2 \varphi)$ modulation of the
azimuthal-dependent $\qt$ cross section $d\sigma/d\qt^2 d\varphi$ behaves as
$1/\qt^2$ at small values of $\qt$.

The production of a generic system $F$ of two or more colourless particles 
(for instance, $F=\{ \gamma \gamma\}, \{ZZ \}, \{ W^+W^-\}$)
is considered in 
Ref.~\cite{Catani:2010pd}. Since the particles in $F$ have no QCD colour
charge, the elastic-production subprocesses (of the type in Eq.~(\ref{lopro}))
that are permitted by colour conservation are $q{\bar q} \to F$ and
$gg \to F$. The study of the small-$\qt$ region performed in 
Ref.~\cite{Catani:2010pd} shows that QCD radiative corrections to the gluon
fusion channel, $gg \to F$, lead to azimuthal correlations with a singular
$\qt$ behaviour (the singular behaviour is instead
absent for QCD corrections to the
quark--antiquark annihilation channel, $q{\bar q} \to F$).
The singular behaviour is proportional to $1/\qt^2$ at the lowest order
(i.e., the corresponding $a_{\rm corr}(\qh)$
of Eq.~(\ref{nlosigma}) is not vanishing) and it is enhanced by 
double-logarithmic factors, $\ln^2(M^2/\qt^2)$, at higher orders \cite{Catani:2010pd}.
Not all the azimuthal harmonics have a singular behaviour in the limit
$\qt \to 0$ (see, in particular, Eqs.~(76)--(86) and accompanying comments in
Ref.~\cite{Catani:2010pd}): the singular harmonics are $\cos (2\varphi)$
(and $\sin (2\varphi)$) starting from the lowest order, and
$\cos (4\varphi)$
(and $\sin (4\varphi)$) at higher orders
(the $\sin (2\varphi)$ and $\sin (4\varphi)$ harmonics typically receive
contribution from parity-violating effects).

The study of Ref.~\cite{Catani:2014qha} considers heavy-quark pair production
($F= \{ Q {\bar Q}\}$) in the small-$\qt$ region. The allowed elastic-production
subprocesses (of the type in Eq.~(\ref{lopro})) are 
$q{\bar q} \to Q {\bar Q}$ and $gg \to Q {\bar Q}$, as in the case of the
colourless systems $F$ considered in Ref.~\cite{Catani:2010pd}.
The important difference with respect to Ref.~\cite{Catani:2010pd}
is that the produced heavy quarks ($Q$ and ${\bar Q}$) carry QCD colour charge,
so that they are sources of additional QCD radiation and produce additional
dynamical effects (both final-state effects and initial/final-state quantum
interference effects). Singular azimuthal correlations, analogous to those in
Ref.~\cite{Catani:2010pd}, are produced by QCD radiative corrections to the
gluon fusion channel $gg \to Q {\bar Q}$. Additional azimuthal correlations,
with the same small-$\qt$ singular behaviour, are produced by the non-vanishing
colour charges of $Q$ and ${\bar Q}$, and they are generated through QCD
radiative corrections to {\em both} production channels 
$q{\bar q} \to Q {\bar Q}$ and $gg \to Q {\bar Q}$. The explicit lowest-order
(which corresponds to $a_{\rm corr}(\qh)$
in Eq.~(\ref{nlosigma})) and higher-order results of Ref.~\cite{Catani:2014qha}
show that the small-$\qt$ singular behaviour of these additional
azimuthal correlations affects the $\cos (n\varphi)$ harmonics with {\em
arbitrary even} values of $n$ ($n=2,4,6,\dots)$.

The analyses and results of Refs.~\cite{Catani:2010pd, Catani:2014qha} involve
some process-dependent features. However, the sources of singular azimuthal
correlations that are identified therein have a process-independent (universal)
dynamical origin. This allows us to draw some general conclusions for the entire
class of processes discussed in Sects.~\ref{sec:corr} and \ref{sec:qt}
(some of these processes are mentioned in Eq.~(\ref{div})).
There are certainly two sources of small-$\qt$ singular azimuthal correlations
and related divergent azimuthal asymmetries for these processes. These two
sources are
\begin{itemize}
\item[ $i)$ ]  initial-state collinear radiation from gluonic colliding states
that produce the particles of the high-mass system $F$;
\item[ $ii)$ ] soft wide-angle radiation (both final-state contributions and
initial/final-state interference contributions) generated by colour-charged
particles of the high-mass system $F$.
\end{itemize}
The process-independent origin of singular azimuthal correlations that we have
just specified leads to the conditions (\ref{a1}) and (\ref{a2}) that we have
anticipated in Sect.~\ref{sec:corr}. Specifically, the collinear origin at the
point $i)$ leads to the condition (\ref{a1}), and the soft origin at the point
$ii)$ leads to the condition (\ref{a2}). Some main consequences of the conditions
(\ref{a1}) and (\ref{a2}) are already discussed in the subsequent part 
of Sect.~\ref{sec:corr}, and they are not repeated here. In the following we
only present some additional comments.

As for the initial-state collinear radiation at the point $i)$, we refer the
reader to Eqs.~(33)--(43) (and accompanying comments) 
in Ref.~\cite{Catani:2010pd} for a more detailed discussion of its
process-independent dynamical origin. Collinear radiation from the initial-state
colliding partons ($c_1$ and $c_2$ in Eq.~(\ref{lopro})) directly affects the
kinematics ($\qt$ and rapidity) of the entire final-state system $F$, but it has
no direct kinematical effects on the azimuthal dependence of the individual
particles in $F$. The azimuthal dependence requires variations of the angular
momentum of those particles. Therefore, collinear radiation can dynamically
produce azimuthal correlations only through spin correlations induced by the
collinear-emission process. Gluonic collinear splitting processes are indeed
intrinsically spin polarized. The produced {\em linear} polarization of the
exchanged ($t$-channel) gluons leads to quantum interferences (between 
scattering amplitudes and their complex-conjugate amplitudes) that generate
$n=2$ (single helicity-flip contributions 
\cite{Nadolsky:2007ba, Catani:2010pd}) and $n=4$ (double 
helicity-flip contributions \cite{Catani:2010pd}) azimuthal harmonics.
The singular azimuthal correlations are due to the initial-state partonic
splitting subprocess $a \to g + X$ ($a=q,{\bar q},g$ is spin unpolarized
and $X$ denotes the
accompanying collinear radiation) that dynamically generates a
linearly-polarized colliding gluon.
Singular azimuthal correlations are analogously produced by the (QED induced)
collinear splitting subprocess $a \to \gamma + X$, which generates a
linearly-polarized colliding photon.
In contrast, quark (antiquark) collinear splitting processes 
($a \to q ({\bar q}) + X$) 
turn out to be spin
unpolarized, as a consequence of helicity conservation in QCD radiation from a
massless quark (antiquark). This difference between the collinear evolution of
gluons (or photons) and quarks (antiquarks) is responsible for the specification of gluon
(or photon) initial states at the point $i)$ and in the condition (\ref{a1}).
Obviously, QCD radiation from quarks can also produce spin correlations (and
ensuing azimuthal asymmetries), as in the case of the DY process, but they are
not singular in the collinear limit and, consequently, in the small-$\qt$
region.

The soft radiation at the point $ii)$ refers to radiation of soft partons at
wide angles with respect to the directions of the colliding partons and of the
particles of the system $F$. If the system $F$ contains only colourless
particles, primary soft radiation only originates from the colliding partons 
$c_1$ and $c_2$ in Eq.~(\ref{lopro}) ($c_1c_2=q{\bar q}$ or $c_1c_2=gg$, in this
colourless case): this radiation is helicity conserving and it has a high degree
of process independence \cite{Catani:2010pd}. If $F$ contains colour-charged
particles \cite{Catani:2014qha}, 
they can radiate wide-angle gluons that kinematically produce 
azimuthal-correlation dependence. In these processes, we can have
$c_1c_2=q{\bar q}, gg, qg$ and so forth, depending on the particles in $F$.
Colour-charged particles in
$F$ and the colliding partons $c_1$ and $c_2$ simultaneously act as sources of primary QCD
radiation, and the dynamical pattern of soft wide-angle radiation has a very
high degree of process dependence. The process dependence originates by colour
correlations with the underlying hard-scattering subprocesses.
Moreover, within a single process, soft wide-angle radiation strongly depends
on the kinematics of the particles in $F$ because of initial/final-state
interferences and related colour-coherence phenomena. All these features have a
`minimal' dependence on the actual partonic content of the colliding hadrons in
the initial state.

We note that the collinear radiation and the soft radiation at the points $i)$
and $ii)$ have a different origin. However, starting from the second non-trivial
perturbative order ($F$ plus two or more additional partons in the final state),
collinear radiation and soft radiation do not act as independent sources of
singular azimuthal correlations. The two sources are indeed entangled by
helicity and colour correlations, and they are analogously dynamically
tangled with the
helicity and colour structure of the underlying hard-scattering
subprocesses (see, e.g., Eqs.~(22)--(25) and accompanying comments in
Ref.~\cite{Catani:2014qha}).

Soft wide-angle radiation generally leads to divergent azimuthal
harmonics with even
values of $n$ \cite{Catani:2014qha}. In principle, we have no reasons
to justify the absence of divergent $n$-odd harmonics that can be produced by 
soft wide-angle radiation.
In the case of $Q{\bar Q}$ production we
cannot exclude the appearance of divergent $n$-odd harmonics from subdominant
logarithmic contributions at high orders. We find it really difficult to justify
the absence of divergent $n$-odd harmonics for processes with a more involved
kinematical structure, such as those processes related to jet production
(e.g., $F=\{V j \}$ or $F=\{ jj \}$). In the case of jets, the more involved
kinematical structure arises, for instance, from jet selection cuts, and it
certainly arises from the (infrared and collinear safe) procedure that is
necessary to identify (define) hard-scattering jets.

To our knowledge, azimuthal correlations for jet production processes have not yet been studied
throughout computations in QCD perturbation theory. In view of our comments on
divergent harmonics with odd values of $n$, in the following we present an
illustrative numerical investigation at the lowest perturbative order
for the process of associated production of a vector boson $V$ 
($V=Z,W^\pm,\gamma^*$
and one jet $j$.

We consider the inclusive production process $h_1+h_2 \to Vj +X$.
The corresponding LO partonic processes are $c_1c_2 \to Vj$ 
(see Eq.~(\ref{lopro})), where $c_1c_2=q{\bar q}, qg, {\bar q}g$ and
$j$ is a single final-state parton (with the corresponding flavour assignment,
namely, $g,q,{\bar q}$). The lowest-order contribution to azimuthal correlations
is entirely due to all the NLO tree-level processes whose final state is 
`$V+2$\,partons',
where (after application of an infrared and collinear safe jet algorithm)
one of the final-state partons is identified with the observed jet while
the other parton contributes to the inclusive final-state radiation.
As we have already discussed in Sect.~\ref{sec:corr}, both conditions (\ref{a1})
and (\ref{a2}) are fulfilled: therefore, at the lowest perturbative order we
expect singular azimuthal correlations of both collinear (see point $i)$ above)
and soft (see point $ii)$ above) origins. Note, however, that the collinear
contribution can lead to singularities only for the $\cos(2 \varphi)$ harmonic,
since a singular $\cos(4 \varphi)$ harmonic requires \cite{Catani:2010pd}
a $gg$ initial state at the LO and a final state with $Vj$ and at least two
additional partons.
The soft contribution is expected to be
responsible for singular $\cos(n \varphi)$ harmonics with all even values of 
$n$ ($n=2,4,6,\dots)$ \cite{Catani:2014qha} and, possibly, also with odd values
of $n$.

We perform the lowest-order calculation of azimuthal harmonics by specifically
considering the case of associated $Zj$ production in $pp$ collisions at the LHC
(${\sqrt s}=8$~TeV). We use the numerical Monte
Carlo program {\tt DYNNLO} \cite{Catani:2009sm} with the same set-up and
parameters 
as mentioned and used in
Sect.~\ref{sec:exa}.
We compute the azimuthal harmonics at fixed values of $\qt$ of the $Zj$
system by integrating over all the other kinematical variables with the 
following
selection cuts on the rapidity\footnote{Rapidity and pseudorapidity are 
equal at this perturbative
order since the jet is formed by a single massless parton.} 
$\eta_{jet}$ and transverse momentum $p_{T jet}$
of the jet $j$: $|\eta_{jet}| < 4.5$ and $p_{T jet} > 20$~GeV.
We do not apply any lower limit on the invariant mass $M$ of the $Zj$ system
since (as stated in Sect.~\ref{sec:exa}) we consider on-shell $Z$ production
and, consequently, $M > M_Z$. 
The jets are defined by using the $k_\perp$~algorithm \cite{Catani:1993hr} or,
equivalently, the anti-$k_\perp$~algorithm \cite{Cacciari:2008gp}, 
since both algorithms give
exactly the same results at this perturbative order. 
We use the value $R=0.4$ of the jet radius 
$R$. We consider 
fully-inclusive $Vj$ production and, therefore, we do not select the observed
jet $j$ according to the hardness of its transverse momentum (the jet $j$ of the
$Zj$ system is in turn one of the two partons of the final state 
`$V+2$\,partons', independently of the value of its transverse momentum).
The conditions that we have just specified give
an infrared and collinear safe definition of the 
cross section (and of related azimuthal correlations)
for inclusive $Zj$ production.
 
We define the azimuthal-correlation angle as 
$\varphi = \phi({\bom p_T}_{jet}) - \phi(\bqt)$, where
$\phi({\bom p_T}_{jet})$ and  $\phi(\bqt)$ are the azimuthal angles of
the transverse momenta ${\bom p_T}_{jet}$ and $\bqt$ of the jet $j$ and the
system $Zj$ in the centre--of--mass
frame of the collision. As discussed and emphasized in Sect.~\ref{sec:corr},
this is one possible definition of the relevant azimuthal-correlation angle
and the singular behaviour of azimuthal correlations is not affected by this
choice. Note that we are not using the azimuthal-angle definition in the CS
frame of the $Zj$ system. Our choice of $\varphi$ has a twofold purpose.
Firstly, we take the chance of showing results for an alternative (with respect
to the CS frame) definition of $\varphi$. Secondly, the definition of the CS
rest frame of the $Zj$ system requires the determination of the {\em
four}-momentum of the jet (e.g., also the jet invariant mass matters) and such
quantity is not always actually measured in experiments
(anyhow, its determination is not necessary by using 
$\varphi = \phi({\bom p_T}_{jet}) - \phi(\bqt)$).

\begin{figure}[th]
\centering
\hspace*{-0.3cm}
\includegraphics[width=4in]{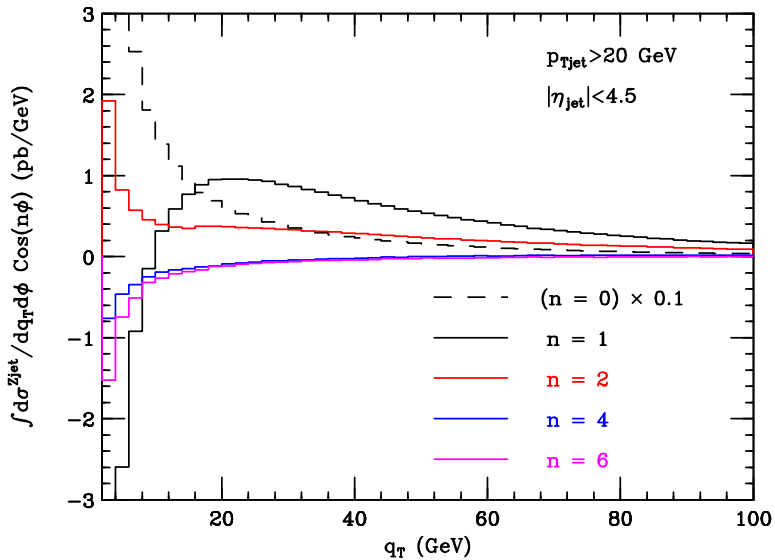}
\caption{
\label{fig:zj}
{\em Lowest-order transverse-momentum spectrum $d\sigma_n/d\qt$ of $cos(n\varphi)$ harmonics in the case of $Z$+jet production.}}
\end{figure}

We consider the $\qt$ spectrum
$d\sigma_n/d\qt$ of $\cos(n\varphi)$ harmonics (see Eq.~(\ref{nqt})), 
and in Fig.~\ref{fig:zj}
we present our lowest-order numerical results for 
$Zj$ production.
Note that in Fig.~\ref{fig:zj} we present the $\qt$ dependence of 
$d\sigma_n/d\qt$, 
whereas the
azimuthal asymmetries in Fig.~\ref{dyfig} are presented as functions of 
$r_{\rm cut}=q_{\rm cut}/M$ and are obtained by integration  
of $d\sigma_n/d\qt$ over $\qt$ with the lower limit $\qt > q_{\rm cut}$.
Note also that we are not showing results
for $d\sigma_n/d\qt^2$
(see Eqs.~(\ref{nlosigma}) and (\ref{nqt})), which is the $\qt$ 
differential cross section that we
typically use throughout the text of this paper. 
In particular, since $d\sigma_n/d\qt= 2 \qt d\sigma_n/d\qt^2$,
the presence of
singular harmonics (with $n=1,2,\dots$) 
implies that $d\sigma_n/d\qt \propto 1/\qt$
at the lowest perturbative order.

We comment on the results in Fig.~\ref{fig:zj}. 
As in the case of Fig.~\ref{dyfig}, in Fig.~\ref{fig:zj} we also present the result
for the harmonic with $n=0$ (i.e., the azimuthally-integrated $\qt$ cross section).
Since we are considering finite values of $\qt$ ($\qt > 2$~GeV), the
plus-prescription (see Eq.~(\ref{nlosigma})) is not effective and the harmonic with
$n=0$ has a dominant behaviour that is proportional to $1/\qt \ln(M/\qt)$
at small values of $\qt$: the corresponding numerical results in 
Fig.~\ref{fig:zj} are consistent with such 
behaviour.
The other results in Fig.~\ref{fig:zj} regard the $n$-th harmonics with the
specific values $n=1,2,4,6$. The numerical results for these azimuthal-correlation
harmonics have similar quantitative features: 
$d\sigma_n/d\qt$ is small 
and it has a mild dependence on $\qt$ 
in the region where
$\qt \gtap 20$~GeV; at smaller values of $\qt$, the $\qt$ shape changes, and the
absolute magnitude of $d\sigma_n/d\qt$ increases rapidly as $\qt$ decreases.
This small-$\qt$ numerical behaviour is consistent with the dependence 
$d\sigma_n/d\qt \propto 1/\qt$, and we interpret these results as a 
clear indication of singular azimuthal harmonics with $n=1,2,4,6$.
As we have previously discussed, the harmonic with $n=2$ has singularities of both
collinear and soft origins. The initial-state collinear radiation at the  
point $i)$
cannot give a singular contribution to the harmonic with $n=1,4,6$: in the case of
these harmonics the singularity is due to the soft wide-angle radiation
at the point $ii)$ (radiation at large angles with respect to the directions
of the initial-state colliding partons and to the direction of the momentum of the
jet $j$ in the $Zj$ system).
If $d\sigma_n/d\qt \propto 1/\qt$ at small values of $\qt$, 
the size of the $\qt$ slope of $d\sigma_n/d\qt$
roughly corresponds to the
size of the $n$-th harmonic of the coefficient $a_{\rm corr}(\qh)$ in
Eq.~(\ref{nlosigma}). From the comparison of
the results with $n=1,2,4,6$ in the small-$\qt$ region of Fig.~\ref{fig:zj},
we notice that the $\qt$ slope of $d\sigma_n/d\qt$ with $n=1$ is sizeable and
this is a clear evidence of singular harmonics with {\em odd}
values of $n$.

\subsection{Resummation}
\label{sec:resaz}

The perturbative resummation of singular azimuthal-correlation contributions
was considered in Refs.~\cite{Catani:2010pd, Catani:2014qha}.
Reference \cite{Catani:2010pd} presents an all-order treatment (i.e., the
treatment is valid to arbitrary logarithmic accuracy) of azimuthal correlations,
but it is limited to processes with colourless systems $F$ (the individual
particles in $F$ have no QCD colour charge). Therefore, 
Ref.~\cite{Catani:2010pd} only deals with azimuthal correlations due to
initial-state collinear radiation (see the point $i)$ previously mentioned in
Sect.~\ref{sec:azor}) for production processes that occur through the gluon
fusion channel ($gg \to F+X$).  The all-order resummation for $Q{\bar Q}$
production in Ref.~\cite{Catani:2014qha} is limited to an explicit treatment up
to NNLL accuracy, but it deals with a specific example of the relevant case in
which the produced final-state system $F$ is colourful (the individual particles
in $F$ carry QCD colour charge). Therefore, the analysis and results of
Ref.~\cite{Catani:2014qha} not only deal with collinear radiation, but they also
deal with all the main conceptual (and technical) aspects related to
azimuthal-correlation effects due to soft wide-angle radiation
(see the point $ii)$ previously mentioned in
Sect.~\ref{sec:azor}). The all-order resummation structure in 
Ref.~\cite{Catani:2014qha} is quite different from that of 
azimuthally-insensitive $\qt$ cross sections (e.g., the case of the DY process
\cite{Collins:1984kg}): the former includes resummation factors and
helicity/colour correlations that originate from the collinear/soft contributions
at the points $i)$ and $ii)$ in Sect.~\ref{sec:azor}. The resummation results
of Ref.~\cite{Catani:2014qha} certainly have some process-dependent features
(due to the specific production of $F=\{ Q{\bar Q} \}$), but they are
generalizable (although with specific differences and, possibly, complications
at the technical level) to generic production processes of the type in
Sect.~\ref{sec:qt} with final-state colourful systems $F$ (e.g., $F=\{ V j\},\{j
j \}$). Therefore, we can consider the all-order resummation for $Q{\bar Q}$
production in Ref.~\cite{Catani:2014qha} as the prototype of resummation for this
entire class of processes and, consequently, we can present some general
comments (by avoiding cumbersome technicalities) in the following.

The all-order resummation in Ref.~\cite{Catani:2014qha} is performed in $\bf b$
space, by treating azimuthally-independent and azimuthal-correlations
contributions on equal footing. Then the $\qt$ cross section is obtained by
inverse Fourier transformation of the resummed result from $\bf b$ space
to $\bqt$ space. To facilitate the discussion of the comparison (analogies and
differences) with the azimuthally-insensitive case (see Sect.~\ref{sec:azav}),
it is convenient to consider the $\qt$ dependence of the $n$-th harmonic: 
\beq
\label{nqt}
\frac{d\sigma_n}{dM^2 d\qt^2} \equiv \int_0^{2\pi} d\varphi \;\cos(n\varphi) 
 \;\frac{d\sigma}{dM^2 \,d\qt^2 \,d\varphi}
 \;\;. 
\eeq
The azimuthal asymmetry in Eq.~(\ref{nh}) is obtained by integration of
Eq.~(\ref{nqt}) over $\qt$.
Considering harmonics, the inverse Fourier transformation from $\bf b$ space
to $\bqt$ space can be recast in the form of a Bessel transformation.
In particular, the $n$-th harmonic projects out the $n$th-order Bessel function
$J_n(x)$ (see, e.g., Sect.~6 in Ref.~\cite{Catani:2010pd}), and 
the final result of the resummation procedure 
of the singular azimuthal-correlation terms  
has the following (sketchy) form:
\beq
\label{resn}
\frac{d\sigma^{(\rm res)}_n}{dM^2 d\qt^2} =
\int_0^{+\infty} db \; b \;J_n(b\qt) \;\Sigma_{n}^{(\rm res)}(M,b) \;\;.
\eeq

The resummation formula (\ref{resn}) allows us to discuss analogies and
differences with respect to the corresponding formula (\ref{resav})
for the azimuthally-independent case (which coincides with the $n=0$ harmonic).
Obviously, we can also consider the $n$-th harmonic that is obtained
through the replacement $\cos(n\varphi) \to \sin(n\varphi)$ of the weight
function in Eq.~(\ref{nqt}). Such a replacement does not change the Bessel
transformation structure of Eq.~(\ref{resn}), although the factor 
$\Sigma_{n}^{(\rm res)}$ has different explicit expressions for cosine and sine
harmonics.
The factor $\Sigma_{n}^{(\rm res)}(M,b)$ in Eq.~(\ref{resn}) embodies the effect
of the all-order resummation procedure in $b$ space. In the following, we first
comment on the perturbative expansion of Eq.~(\ref{resn}) and, then, we discuss
the effect of perturbative resummation.

At the lowest perturbative order,
the factor $\Sigma_{n}^{(\rm res)}(M,b)$ in the integrand of Eq.~(\ref{resn})
corresponds to the Fourier transformation of the singular contribution
$a_{\rm corr}(\qh)/\qt^2$ in Eq.~(\ref{nlosigma}).
Therefore, $\Sigma_{n}^{(\rm res)}(M,b)$ is simply proportional to the $n$-th
harmonic of the azimuthal function $a_{\rm corr}(\qh)$ and its dependence on $b$
is only due to the Bessel transformation of $1/\qt^2$. This $b$ dependence is
that of the following integral over $\qt$:
\beq
\label{i0n}
{\widehat I}^{\;[\,n\,]}_0 (b) = \int_0^{+\infty} \frac{d\qt^2}{\qt^2} \;J_n(b\qt) 
= \frac{2}{n}  \;\;,
\quad \quad (n=1,2,3 \dots) \;\;,
\eeq
We note that, having selected the $n$-th harmonic, the non-integrable 
singular function 
$1/\qt^2$ has a well defined Bessel transformation. We also note that this 
Bessel transformation is a constant function of $b$ in the large-$b$ 
region\footnote{We can
introduce an upper bound $\qt \ltap M$ in the integral of Eq.~(\ref{i0n}),
as it would be appropriate to deal with the small-$\qt$ region (approximation).
In the large-$b$ limit ($bM \gg 1$), the result of such integral is that in the
right-hand side of Eq.~(\ref{i0n}), modulo additive vanishing corrections
(corrections that vanish as powers of $1/(bM)$).}.

At higher perturbative orders, singular azimuthal correlations produce a
dependence of 
the $\qt$ cross section  
$d\sigma/dM^2 d^2\bqt$
on $1/\qt^2$ times powers of $\ln(M^2/\qt^2)$
(with at most two powers of
$\ln(M^2/\qt^2)$ for each additional power of $\as$).
As discussed in Sect.~\ref{sec:qt}, such $\qt$ dependence is 
factorized
with respect to the dependence on the azimuthal-correlation angle.
Therefore, the $b$ dependence of $\Sigma_{n}^{(\rm res)}(M,b)$ is due to the 
Bessel transformation of these singular functions of $\qt$.
For example, in the case of the 2nd harmonic the dominant (DL) next-order
correction is proportional to the following $\qt$ space integral:
\beq
\label{i22}
{\widehat I}^{\;[\,2\,]}_2 (M,b) = \int_0^{+\infty} 
\frac{d\qt^2}{\qt^2} \;\ln^2\!\left(\frac{M^2}{\qt^2}\right)
\,J_2(b\qt) =
\;\ln^2\!\left(\frac{b^2M^2}{b_0^2}\right)
- 2 \ln\!\left(\frac{b^2M^2}{b_0^2}\right) + 2 \;\;,
\eeq
where $b_0=2 e^{-\gamma_E}$ ($\gamma_E=0.5772\dots$ is the Euler number)
is a numerical coefficient (of kinematical origin) that
customarily appears in the context of Fourier (Bessel) transformations of
logarithmic functions.
In general, the typical singular behaviour of $d\sigma_n/dM^2 d\qt^2$ 
at small $\qt$ leads to contributions to $\Sigma_{n}^{(\rm res)}(M,b)$
that are proportional to the following basic integrals:
\beq
\label{ink}
{\widehat I}^{\;[\,n\,]}_k (M,b) \equiv \int_0^{+\infty} 
\frac{d\qt^2}{\qt^2} \;\ln^k\!\left(\frac{M^2}{\qt^2}\right)
\,J_n(b\qt) \;\;,
\quad \quad (n=1,2,3 \dots)
 \;\;.
\eeq
The result of the basic integrals is
\beq
\label{inkres}
{\widehat I}^{\;[\,n\,]}_k (M,b) = \frac{2}{n}
\ln^k\!\left(\frac{b^2M^2}{b_0^2}\right)
+ \sum_{m=1}^k \frac{k! \; (-1)^m}{m! \, (k-m)!} \;{\widehat d}^{\;[\,n\,]}_m 
\;\ln^{k-m}\!\left(\frac{b^2M^2}{b_0^2}\right)  \;,
\quad \!\!(k \geq 1, n=1,2,3,\dots) \,,
\eeq
where ${\widehat d}^{\;[\,n\,]}_m$ are numerical coefficients that depend 
on $n$ and $m$ (see Eq.~(\ref{dnm})).
Note that each of these integrals ${\widehat I}^{\;[\,n\,]}_k (M,b)$ is 
a polynomial of degree $k$ in the variable $\ln(bM)$.

We note that the result in Eq.~(\ref{inkres}) for the integrals 
${\widehat I}^{\;[\,n\,]}_k$
can be obtained by using the generating function method that is used in
Appendix~B of Ref.~\cite{Bozzi:2005wk}. The relevant generating function is
\beq
\label{gf}
\int_0^{+\infty} \frac{d\qt^2}{\qt^2} \;\left(\frac{\qt^2}{M^2}\right)^{\!\lambda}
 \;J_n(b\qt) = 
\;\left(\frac{b_0^2}{b^2M^2}\right)^{\!\lambda}
\; {\widehat d}^{\;[\,n\,]}(\lambda) \;\;,
\eeq
with
\beq
\label{gfd}
{\widehat d}^{\;[\,n\,]}(\lambda) = \left(\frac{2}{b_0}\right)^{\!2\lambda}
\; \frac{\Gamma(\lambda+ n/2)}{\Gamma(1 - \lambda + n/2)} \;\;,
\eeq
where $\Gamma(z)$ is the Euler Gamma function.
In particular, the coefficient ${\widehat d}^{\;[\,n\,]}_m$ in Eq.~(\ref{inkres})
is the $m$-th derivative of ${\widehat d}^{\;[\,n\,]}(\lambda)$ with respect 
to $\lambda$ at the point $\lambda=0$:
\beq
\label{dnm}
{\widehat d}^{\;[\,n\,]}_m = \left[ \left( \frac{d}{d\lambda} \right)^{\!m}
\;{\widehat d}^{\;[\,n\,]}(\lambda)
 \right]_{\lambda=0} \;\;.
\eeq

We remark again that the singular (and non-integrable) azimuthal-correlation
contributions in $\qt$ space at each f.o. have a corresponding $n$th-order
Bessel transformation in $b$ space
(see Eqs.~(\ref{i0n})--(\ref{inkres})).
Since the singular terms have an azimuthal dependence that is factorized with
respect to the $\qt$ dependence in $\qt$ space, our remark equally applies
to the two-dimensional Fourier transformation from $\bqt$ space to ${\bf b}$
space of the entire azimuthal-correlation component 
$d\sigma^{\rm corr}/dM^2 d^2\bqt$ of the cross section
(the projection onto the $n$-th harmonic in $\bqt$ space and the sum over an
infinite set of harmonics in ${\bf b}$ space are procedures that can be applied
with no further conceptual difficulties).
The general f.o. structure of $\Sigma_{n}^{(\rm res)}(M,b)$ in the large-$b$
region is formally similar to that of the corresponding function 
$\Sigma_{\rm az. av.}^{(\rm res)}(M,b)$ for the azimuthally-averaged
cross section. In both cases we are dealing with logarithmically-enhanced
singular terms in $\qt$ space at small values of $\qt$ ($\qt \ll M$)
and with corresponding powers of $\ln(bM)$ and constant terms in $b$ space
at large values of $b$ ($bM \gg 1$). In both cases the all-order resummation
procedure in $b$ space deals with the resummation of $\ln(bM)$ terms, and 
the formally dominant contributions at each f.o. are of DL type
(two additional powers of $\ln(bM)$ for each additional power of $\as$).
In the case of azimuthal correlations, the $\ln(bM)$ and constant terms
originate from the Fourier (Bessel) transformation of non-integrable functions 
of $\qt$ (see Eq.~(\ref{ink})). This origin can be contrasted at the formal level
with that
of the analogous contributions
to the azimuthally-averaged case, in which the  $\ln(bM)$ (and constant) terms
arise from the Fourier (Bessel) transformation of terms that are
plus-distributions (and $\delta$-function contributions) of $\qt$.
The basic integrals that occur in the azimuthally-averaged case are obtained
from the right-hand side of
Eq.~(\ref{ink}) throughout the replacement $J_n \to J_0$ and the
replacement of the $\qt$ functions with the corresponding plus-distributions.
The explicit computation of these integrals
for the azimuthally-averaged case is presented in the Appendix~B of 
Ref.~\cite{Bozzi:2005wk} (see, in particular, Eqs.~(129) and (141) of the arXiv
version of Ref.~\cite{Bozzi:2005wk} or, equivalently, Eqs.~(B.18) and (B.30)
of the corresponding published version).

We now discuss the all-order behaviour of the resummed result in 
Eq.~(\ref{resn}). The $b$-space resummed cross section $\Sigma_{n}^{(\rm res)}$
involves several factors (see Sect.~6 in Ref.~\cite{Catani:2010pd}
and Sect.~2 in Ref.~\cite{Catani:2014qha}) and, in particular, the 
explicit dependence on the $n$-th harmonic is definitely related to the specific
multiparticle system $F$ that is produced in the hard-scattering process.
Although the resummed functions $\Sigma_{\rm az. av.}^{(\rm res)}$ and
$\Sigma_{n}^{(\rm res)}$ in Eqs.~(\ref{resav}) and (\ref{resn}) are different,
the analysis in Refs.~\cite{Catani:2010pd, Catani:2014qha} shows that the
dominant DL behaviour of these two $b$-space cross sections is controlled by the
same universal (process-independent) factor, namely, the Sudakov form factor
$S_c(M,b)$ that we have already recalled in Sect.~\ref{sec:azav}.
The presence and universality of this common factor, $S_c(M,b)$, in the resummed
result
of both cross sections in Eqs.~(\ref{resav}) and (\ref{resn}) 
follow from the physical origin of the DL contributions to both cross sections:
the DL terms originate from radiated partons that are soft and collinear to the
initial-state colliding partons $c_1$ and $c_2$ of the corresponding main
production channels
(see Eq.~(\ref{lopro})) in the small-$\qt$ region.

As in the case of azimuthally-independent contributions (see our comments in
Sect.~\ref{sec:azav}),
the Sudakov form factor produces a strong suppression of the large-$b$ region
($bM \gg 1$) in the integrand on the right-hand side of Eq.~(\ref{resn}).
This damping effect leads to resummed perturbative predictions for the $\qt$
dependence of the $n$-th azimuthal harmonic $d\sigma_n/dM^2 d\qt^2$
that have a smooth and, especially, {\em integrable} behaviour in the small-$\qt$
region. The integrable behaviour can be checked by performing the limit 
$\qt \to 0$ of Eq.~(\ref{resn}). Owing to the Sudakov-type suppression of the
large-$b$ integration region, we can proceed (as in Sect.~\ref{sec:azav}) to use
the small-$\qt$ approximation $J_n(b\qt) = {\cal O}((b\qt)^n)$ in the right-hand
side of Eq.~(\ref{resn}), and we obtain \cite{Catani:2010pd}
\beq
\label{avqtn}
\frac{d\sigma^{(\rm res)}_{n}}{dM^2 d\qt^2} \propto \qt^n \;\;
\quad \quad \quad (\qt \to 0) \;\;.
\eeq 
 
The small-$\qt$ behaviour in Eq.~(\ref{avqtn}) is integrable at $\qt=0$ for every
harmonic $(n=1,2,3,\dots)$. This is a highly non-trivial result of the all-order
resummation procedure of terms that individually behave in a very singular way as
$1/\qt^2$ (modulo powers of $\ln(M^2/\qt^2$)).
In the case of the azimuthally-averaged cross section, as the result of the
resummation procedure, the fixed-order singular behaviour is turned into the
constant behaviour of Eq.~(\ref{avqt0}). In the case of the $n$-th azimuthal
harmonic the effect of resummation is even more substantial, since the 
$1/\qt^2$ singular behaviour is turned into the suppressed power-like behaviour,
$\qt^n$, of Eq.~(\ref{avqtn}). We thus expect that the shape of the resummed
$\qt$ spectrum, $d\sigma_n/dM^2 d\qt^2$, for the $n$-th azimuthal harmonic can be
substantially different from the shape of the $\qt$ spectrum for the
azimuthally-averaged cross section. Independently of the detailed $\qt$ shape,
the all-order perturbative resummation has a `dramatic' effect on the total
production rate of the azimuthal asymmetries. The total
($\qt$ integrated) azimuthal asymmetries turn out to be finite (and computable)
after resummation, whereas they can be divergent at fixed perturbative orders.
In contrast, the total ($\qt$ integrated) azimuthally-averaged cross section 
is finite order-by-order in perturbation theory, and it is basically not affected
by resummation effects in the small-$\qt$ region.
 
As we have discussed in Sect.~\ref{sec:qt}, the f.o. divergences of azimuthal
asymmetries are produced by primary real emission of one final-state parton, 
which
is absolutely not compensated by virtual radiation at the corresponding
perturbative order.
At higher perturbative orders, real emission (which is strictly necessary to
produce azimuthal asymmetries) is accompanied by ensuing virtual and real
radiative effects. Throughout all-order resummation, virtual radiative effects
turn out to be dominant in the small-$\qt$ region, and they produce the Sudakov
form factor suppression that makes the azimuthal asymmetries finite (and with a
smooth $\qt$ shape at small values of $\qt$).

The small-$\qt$ behaviour in Eqs.~(\ref{avqt0}) and (\ref{avqtn}) is the result
of a combined effect of QCD dynamics and kinematics. We conclude this subsection
with a brief discussion of this statement.

\setcounter{footnote}{2}

We note that the $\qt$ dependence in Eqs.~(\ref{avqt0}) and (\ref{avqtn})
is eventually driven by the corresponding small-$\qt$ behaviour of the Bessel
functions $J_n(b\qt)$ in the right-hand side of 
Eqs.~(\ref{resav}) and (\ref{resn}).
The presence of the Bessel function $J_n$ in Eqs.~(\ref{resav}) and (\ref{resn})
has a kinematical origin: the Bessel functions simply arise from the 
{\em two-dimensional} Fourier
transformation from $\bf b$ space to $\bqt$ space after projection onto the
$n$-th harmonic component. This kinematical origin, however, does not imply that
the small-$\qt$ behaviour in Eqs.~(\ref{avqt0}) and (\ref{avqtn}) is purely
produced by (all-order) kinematical effects. Such small-$\qt$ behaviour 
has indeed a dynamical origin, since it is due to the strong Sudakov-type
suppression of the large-$b$ integration region 
(see Eqs.~(\ref{resav}) and (\ref{resn})) that is embodied in the resummed
components $\Sigma_{\rm az. av.}^{(\rm res)}$ and $\Sigma_{n}^{(\rm res)}$. To
make clear the role of this strong suppression, we can consider an explicit
counterexample 
as given by a $b$ space function $\Sigma_{n}(M,b)$ that is
is power-like suppressed by the factor 
$(1/bM)^{2\lambda}$ in the large-$b$ region\footnote{This is exactly the large-$b$
behaviour that is produced by the small-$\qt$ non-singular terms (i.e., the terms
that are simply denoted by the dots on the right-hand side of 
Eq.~(\ref{nlosigma})).} 
(we assume that $\lambda$ is positive, and we may also assume that $\lambda$
is arbitrarily large if we want to obtain a suppression effect that is
quantitatively strong).
The Bessel transformation (see the right-hand side of
Eq.~(\ref{resn})) of such a power-like function leads\footnote{The reader can
easily check this result, for instance, by inspection of the Bessel integral
in Eq.~(\ref{gf}).}
to a $\qt$ differential cross section $d\sigma_n/dM^2 d\qt^2$
that is equally power-like suppressed as $(\qt^2)^{\lambda -1}$ in the small-$\qt$
region, with no power-like dependence on the order $n$ of the Bessel 
function $J_n$.
The Sudakov-type suppression in $b$ space acts differently on the
azimuthally-averaged cross section and on the azimuthal asymmetries in $\qt$
space. In the case of the azimuthally-averaged cross section
$d\sigma_{\rm az.av.}^{({\rm res})}/dM^2 d\qt^2$, resummation leads to a smearing
of the LO $\delta$-function behaviour (see Eq.~(\ref{losigma})), and the resummed
$\qt$ distribution broadens at small values of $\qt$. In the case of the
azimuthal asymmetry $d\sigma_n^{({\rm res})}/dM^2 d\qt^2$, resummation leads to a
suppression of the $1/\qt^2$ behaviour at the lowest order, and the 
resummed $\qt$ distribution is power-like suppressed at small values of $\qt$.
In both cases, all-order kinematical effects have a non-negligible role.
Transverse-momentum conservation, which is `exactly' (provided $\qt \ll M$)
implemented through $b$
space resummation, is eventually responsible for the small-$\qt$ behaviour 
in Eqs.~(\ref{resav}) and (\ref{resn}), which is not a DL exponentially suppressed
$\qt$ behaviour. This feature is (partly) a consequence of the 
{\em two-dimensional}
nature of the kinematical conservation law of the transverse-momentum vector
$\bqt$, which is the relevant kinematical variable for the small-$\qt$ 
behaviour of the azimuthally-averaged cross section and of the azimuthal
asymmetries. In contrast, we can recall the effect of resummation on observables,
such as event shapes, that are controlled by kinematical variables (e.g.,
energies and invariant masses) with one-dimensional conservation laws:
near the exclusive boundary of the phase space, these observables have a DL
exponential suppression (see, e.g., Ref.~\cite{Catani:1992ua}), 
a suppression effect that is
stronger than that produced by
any fixed power of the kinematical distance from the phase
space boundary.

As we have discussed in Sect.~\ref{sec:exa}, QED radiative corrections can also
produce singular azimuthal asymmetries (for instance, in the DY process
$q{\bar q} \to Z(e^+e^-)+X$). The QCD resummation procedure that we have
discussed in this subsection can equally be applied to QED radiative corrections,
and it conceptually leads to the same results and resummation effects that we
have just discussed for the QCD case. The same reasoning and conclusions are
basically valid in a pure QED context, such as, for instance, in the QED
computation of
$e^+e^- \to \mu^+ \mu^- +X$ (although the 
resummed $b$-space expression and the QED Sudakov form factor acquire an
explicit dependence on the mass of the colliding leptons in the initial state).

\subsection{Azimuthal asymmetries in $t{\bar t}$ production}
\label{sec:ttbar}

To illustrate resummed results 
on the $\qt$ dependence of azimuthal asymmetries
at the quantitative level, we choose the specific process of $t{\bar t}$
production. Our choice has several reasons, besides the fact that 
$t{\bar t}$ production is certainly a process of high phenomenological 
relevance.
One reason is that for $t{\bar t}$ production we have explicit theoretical
control \cite{Catani:2014qha} of singular azimuthal correlations and their QCD
resummation. Moreover, due to the non-vanishing QCD colour charge of the produced
$t$ and ${\bar t}$, in this process we are dealing with singular 
azimuthal-correlation effects of both collinear and soft origin
(see the points $i)$ and $ii)$ in Sect.~\ref{sec:azor}).
These effects appear in the QCD radiative corrections to both production 
channels of
quark-antiquark annihilation ($q{\bar q} \to t{\bar t} +X$) and
gluon fusion ($gg \to t{\bar t} +X$). Finally, singular azimuthal correlations
first arise at the NLO level with respect to the computation of the $t{\bar t}$
total
cross section and, therefore, they are
expected to produce effects that have a relatively-large size.

Perturbative resummation of singular azimuthal correlations is under theoretical
control \cite{Catani:2010pd} also in the case of production processes of
colourless systems $F$ (e.g., 
$F\!=\{ \gamma \gamma \}$, $\{Z \gamma\}$, $\{ZZ\}$, $\{W^+W^-\}$).
However, in these cases we are sensitive only to singular contributions of
collinear origin (see the point $i)$ in Sect.~\ref{sec:azor}) and not to those of
soft origin (see the point $ii)$ in Sect.~\ref{sec:azor}). These singular 
contributions of collinear origin arise only from QCD radiative corrections
to the sole gluon fusion production channel $gg \to F + X$.
The colourless system $F$ is typically produced at
the LO throughout the quark-antiquark production channel ($q{\bar q} \to F$).
Owing to the absence of direct coupling of the gluons to the colourless particles
in $F$, the production channel $gg \to F$ is suppressed by some powers of $\as$
with respect to the channel $q{\bar q} \to F$: typically, 
$gg \to F$ enters at the NNLO in the computation of the total cross section, and
the singular azimuthal asymmetries due to $gg \to F+X$ first appear at the
N$^3$LO. 
Therefore, there is a formal mismatch (suppression) of 
{\em two powers}\footnote{This formal conclusion 
(based on counting the powers of $\as$) 
has a caveat, since the channel $gg \to F$ can receive a quantitative
enhancement with respect to the channel $q{\bar q} \to F$ 
from the possibly large luminosity of the gluon PDF. 
In particular, the gluon PDF can be larger than the
antiquark PDF of the proton. Roughly speaking, the gluon PDF enhancement can
quantitatively compensate the effect of a {\em single power} of $\as$
(see, e.g., the results in Ref.~\cite{Catani:2011qz} for the diphoton, 
$F=\{ \gamma \gamma \}$,
production process).}
of $\as$ of the expected relative size of singular azimuthal-correlation effects
(which turn out to be finite after QCD resummation) in going from 
$t{\bar t}$ production to the production of a colourless system $F$. Owing to
all the reasons that we have just discussed, we think it is more `interesting'
to focus our illustrative quantitative analysis on the case of 
$t{\bar t}$ production.

We consider $t{\bar t}$ production in $pp$ collisions with the centre--of--mass
energy ${\sqrt s}=8$~TeV at the LHC. In Sect.~\ref{sec:exa} 
(see Fig.~\ref{dyfig}-right) we have considered the computation of
$\cos(n\varphi)$ harmonics, where $\varphi$ is the azimuthal angle of the top
quark in the CS frame (the differential cross section is integrated 
over the polar angle of the top quark and over the rapidity $y$ and invariant
mass $M$ of the $t{\bar t}$ pair). The $n$-th harmonics 
in Fig.~\ref{dyfig}-right are integrated over the transverse momentum $\qt$ of
the $t{\bar t}$ pair (with the lower limit $\qt > q_{\rm cut}$). In this
subsection we present numerical results on the $\qt$ dependence of the $n$-th
harmonic, and we explicitly consider $d\sigma_n/d\qt$ rather than 
$d\sigma_n/d\qt^2$.
Since $d\sigma_n/d\qt= 2 \qt d\sigma_n/d\qt^2$,
the presence of
singular harmonics implies that $d\sigma_n/d\qt \propto 1/\qt$ (modulo logs) as
$\qt \to 0$ in f.o. perturbative calculations.

\begin{figure}[th]
\centering
\hspace*{-0.3cm}
\subfigure[]{
\includegraphics[width=3.3in]{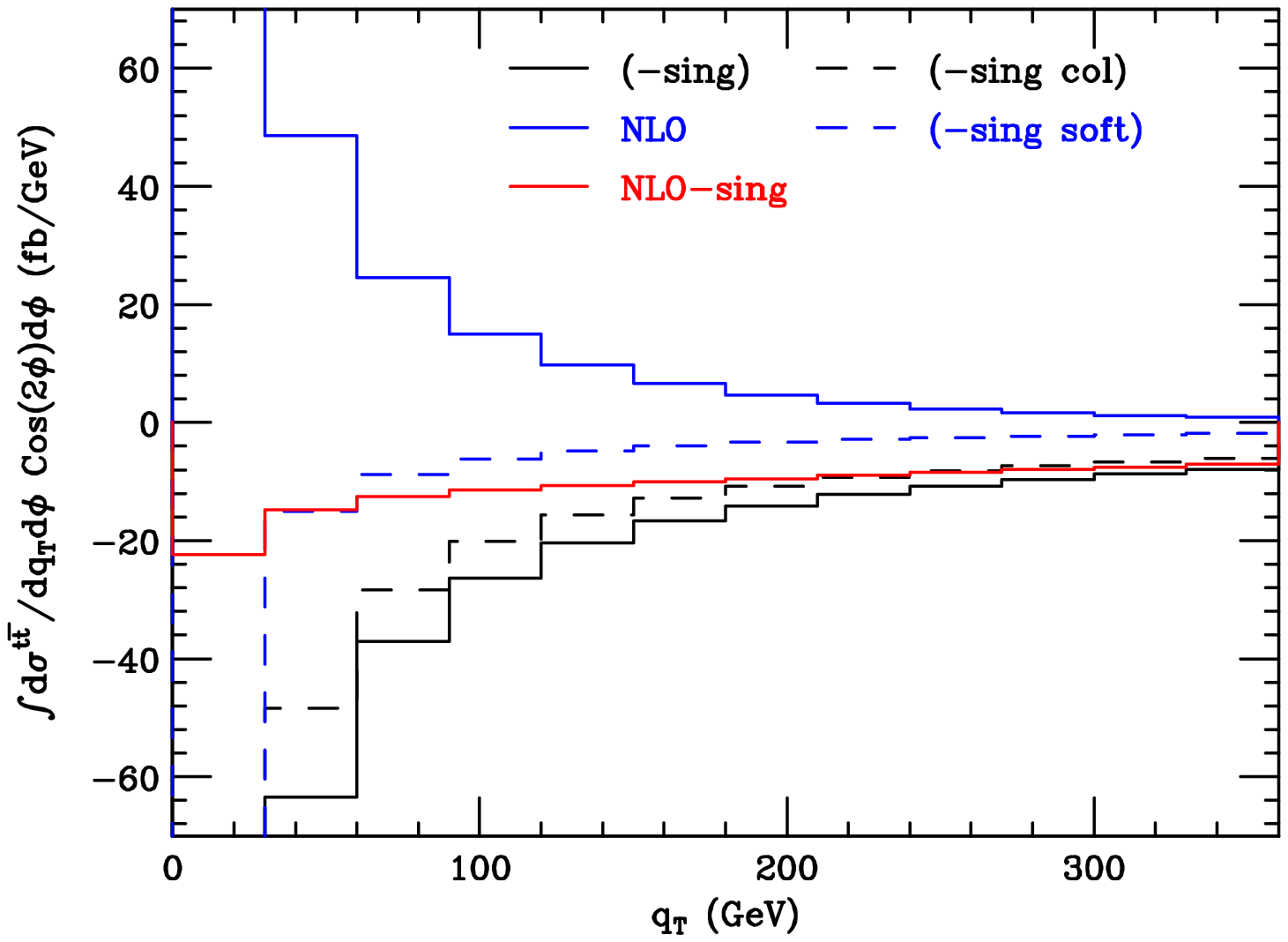}
}
\subfigure[]{
\includegraphics[width=3.3in]{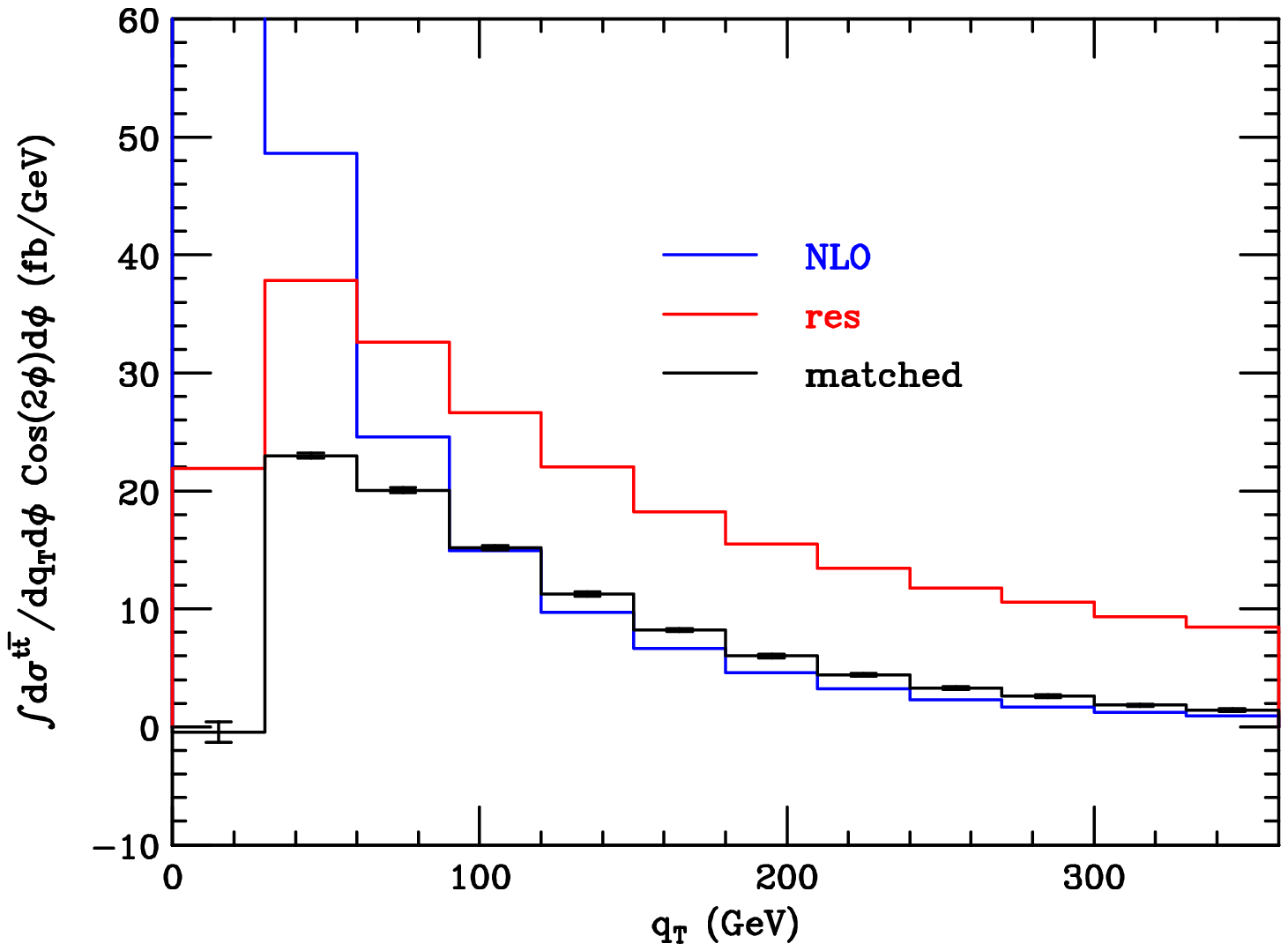}}
\caption{
\label{tt2}
{\em Transverse-momentum spectrum $d\sigma_n/d\qt$ for the $n=2$ harmonic in
$t\bar{t}$ production. In the left panel the exact NLO result (blue) is compared
to the $1/q_T^2$ singular component (black), which is split in its collinear
(black dashed) and soft (blue dashed) parts. The singular components are plotted
with opposite overall sign. The finite result (NLO-sing) that is obtained by
subtracting the singular component from the exact NLO result is also shown (red). In
the right panel the NLO result (blue) is compared to the purely resummed result
(red) and to the complete matched result (black), computed as described in the text.}}
\end{figure}

We specifically consider the $\cos(2\varphi)$ harmonic. We present results at the
lowest perturbative order (Fig.~\ref{tt2}-left)
and including all-order resummation (Fig.~\ref{tt2}-right).

The set-up and parameters of the lowest-order calculation are the same as those
described and used in Sect.~\ref{sec:exa}. Specifically, we compute the
NLO contribution to $t{\bar t}$ production that is due to all tree-level partonic
processes whose final state is `$t{\bar t} + 1$\,parton'.
The result 
of the $\qt$ dependence of the $\cos(2\varphi)$ asymmetry at NLO is
shown in Fig.~\ref{tt2}-left: the cross section increases by decreasing 
$\qt$ and,
in the small-$\qt$ region, the increasing behaviour is consistent with the
$1/\qt$ behaviour predicted in Ref.~\cite{Catani:2014qha}.
The NLO result in the first $\qt$-bin (which includes the region where
$\qt=0$) is not shown in Fig.~\ref{tt2}-left since it would be divergent
to $+\infty$.
The singular $1/\qt$ behaviour predicted in Ref.~\cite{Catani:2014qha} in
analytic form
can be evaluated numerically, and the corresponding result is also presented in 
Fig.~\ref{tt2}-left. Actually, in Fig.~\ref{tt2}-left we plot the NLO singular
result with the opposite overall sign (we do that for better visibility, since
this avoids 
the presentation of close histograms).
We remark that the NLO singular result in Fig.~\ref{tt2}-left {\em exactly}
behaves as $1/\qt$
(i.e., it corresponds to the 2nd harmonic of the term $a_{\rm corr}(\qh)/\qt^2$
in Eq.~(\ref{nlosigma})),
with no additional $\qt$ dependence due to NLO contributions
that are non-singular in the limit $\qt \to 0$ (i.e., the terms denoted by dots
in the right-hand side of Eq.~(\ref{nlosigma})).
In Fig.~\ref{tt2}-left we also present the numerical result (it is denoted by the
label `NLO~$-$~sing') that is obtained by subtracting the $1/\qt$ singular part
from the complete NLO result. Therefore, `NLO~$-$~sing' is expected (predicted)
\cite{Catani:2014qha} to be finite in the limit $\qt \to 0$ and, moreover, 
from it we can numerically extract the behaviour of the dominant non-singular
contribution to the NLO asymmetry in the small-$\qt$ region.
By direct inspection of Fig.~\ref{tt2}-left we see that `NLO~$-$~sing' is indeed
finite: in the limit $\qt \to 0$ the finite result appears to be numerically
consistent with a non-vanishing (actually, negative) constant value. This implies
that, in the case of the $\cos(2\varphi)$ asymmetry, the dominant non-singular
behaviour of $d\sigma_n/d\qt$ is proportional to a 
{\em constant}\footnote{In the case of the DY process, QCD radiative 
corrections do not produce singular harmonics in the limit $\qt \to 0$.
At the lowest perturbative order the small-$\qt$ behaviour of the
$\cos(2\varphi)$ harmonic is $d\sigma_{DY, n=2}/d\qt \propto
\qt \ln(M/\qt)$ \cite{Boer:2006eq} 
(i.e., we have $[d\sigma]_2 \propto \ln(M/\qt)$, 
where 
$[d\sigma]_2$ is the $\qt$ cross section in 
Eq.~(\ref{dyphi})), and the $\cos(2\varphi)$ asymmetry {\em vanishes} 
in the limit
$\qt \to 0$. We note that the dominant non-singular term for the 
$\cos(2\varphi)$ asymmetry in $t{\bar t}$ production is enhanced by a power of
$1/\qt$ (modulo logs) with respect to the DY process.}.

As we have discussed, singular azimuthal correlations can have collinear and soft
origins (see the points $i)$ and $ii)$ in Sect.~\ref{sec:azor}).
In the case of $t{\bar t}$ production, the collinear and soft contributions were
separately evaluated in Ref.~\cite{Catani:2014qha}. In Fig.~\ref{tt2}-left
we report the numerical results of the splitting of the complete NLO singular
contribution in its collinear and soft parts. We note that the collinear and
soft parts have the same relative sign, so that there is no numerical compensation
between them. Both the collinear and soft parts behave as $1/\qt$: removing one
of these two parts from the singular contribution would lead to a divergent
result for the combination `NLO~$-$~sing' in the limit 
$\qt \to 0$.
The overall size of the soft part is smaller than that of the collinear part,
however, the soft part is definitely not negligible: the ratio between the soft
and collinear parts is approximately 0.3.
Moreover, we remark that the relative size of collinear and soft contributions
depends on the type of colliding hadrons and on 
the centre--of--mass energy of the collision\footnote{
This remark follows from the following point. The soft contribution affects both
the gluon fusion ($gg \to t{\bar t}+X$) and $q{\bar q}$ annihilation 
($q{\bar q} \to t{\bar t}+X$) channels, whereas the collinear contribution only
occurs through the gluon fusion channel. These two production channels contribute
to the NLO result with a relative size that depends on the $gg$ and $q{\bar q}$
PDF luminosities. In the case of $t{\bar t}$ production in $pp$ collisions at 
${\sqrt s}=8$~TeV, the $q{\bar q}$ luminosity is much smaller than the $gg$
luminosity. However, by decreasing the relative size of the $gg$ luminosity,
the relative effect of the soft (vs. collinear) contribution would increase.
For instance, the soft/collinear ratio increases in going from $pp$ collisions
to $p{\bar p}$ collisions at the same or smaller (e.g., Tevatron) 
centre--of--mass energy. Similar PDF related considerations about 
the relative
effect of collinear and soft contributions are valid for production processes
of other high-mass systems $F$.}.

We also comment on the results of Fig.~\ref{tt2}-left in the large-$\qt$ region.
We note that `NLO~$-$~sing' is negative, and this is what we can generally expect
at sufficiently large values of $\qt$. The NLO calculation is expected to be
physically well behaved at large $\qt$ and, hence, the NLO result should be
suppressed stronger than $1/\qt$. On the contrary, according to our definition,
the small-$\qt$ singular approximation behaves exactly as $1/\qt$ over the entire
$\qt$ range. As a consequence, at large values of $\qt$
the singular approximation overshoots the NLO
result, and the result of `NLO~$-$~sing' turns out to be negative.

We now move to consider resummation for the $\qt$ dependence of the 
$\cos(2\varphi)$ asymmetry.
Our quantitative results are presented in Fig.~\ref{tt2}-right. 
Before discussing the results in Fig.~\ref{tt2}-right, we briefly comment on the
actual content of our resummed calculation. Details on the calculation will be
presented in a forthcoming paper, in which we also consider resummation results
for the customary
azimuthally-integrated $\qt$ cross section in $t{\bar t}$ production.

We use the resummed theoretical results of Ref.~\cite{Catani:2014qha}
for $t{\bar t}$ production by applying the implementation formalism of
Ref.~\cite{Bozzi:2005wk}. This formalism has so far been applied to
transverse-momentum resummation of azimuthally-integrated (or
azimuthally-insensitive) cross sections for many production processes
(see, e.g., Ref.~\cite{Catani:2015vma} and related references therein).
The complete result $d\sigma$ for the azimuthal-correlation component of the 
$\qt$ differential cross section is obtained by using a matching procedure that
combines the resummation of the singular contributions ($d\sigma^{(\rm res)}$)
with the evaluation of non-singular (finite) contributions
($d\sigma^{(\rm fin)}$) at a given 
f.o. in perturbation theory.
We write $d\sigma=d\sigma^{(\rm res)}+d\sigma^{(\rm fin)}$, and we work at
NLL+NLO accuracy for the resummed (NLL) and f.o. (NLO) contributions.
Therefore, the f.o. part  $d\sigma^{(\rm fin)}$ that contributes to the complete
matched (res + fin) 
result $d\sigma$ in Fig.~\ref{tt2}-right is exactly the 
`NLO~$-$~sing' contribution in Fig.~\ref{tt2}-left.
The resummed part $d\sigma^{(\rm res)}$ is first computed in $b$ space, where we
are dealing with the resummed factor $\Sigma^{(\rm res)}(M,{\bf b})$.
The resummed factor $\Sigma^{(\rm res)}(M,{\bf b})$ 
is proportional to the
contribution from the singular part of the NLO result (see Fig.~\ref{tt2}-left)
and, very roughly speaking, it is multiplied by a form factor
$\exp \{ {\cal G}(M,b) \}$ that is evaluated in resummed form up to NLL accuracy.
We mean that in the large-$b$ region ($bM \gg 1$) we have
${\cal G}(M,b)={\cal G}^{LL}(M,b)+{\cal G}^{NLL}(M,b)$, where 
${\cal G}^{LL}(M,b)$ and ${\cal G}^{NLL}(M,b)$ embody the resummation 
($\sum_{k=1}^{\infty}$) of the LL terms $\as^k \ln^{k+1}(bM)$ and the NLL terms
$\as^k \ln^{k}(bM)$, respectively. 
We remark an important difference between ${\cal G}^{LL}$ and ${\cal G}^{NLL}$.
The LL 
contribution\footnote{If we restrict the evaluation of ${\cal G}^{LL}$ to DL
accuracy (i.e., only $k=1$), $\exp\{{\cal G}^{LL}\}$ is exactly the DL
approximation of the Sudakov form factor that we have mentioned in 
Sects.~\ref{sec:azav}
and \ref{sec:resaz}.}
${\cal G}^{LL}(M,b)$ is essentially process independent and it basically depends 
on the initial-state colliding partons of the partonic 
subprocesses\footnote{In the gluon fusion channel
${\cal G}^{LL}(M,b)$ is the same as the corresponding contribution 
to Higgs boson production \cite{Bozzi:2005wk, deFlorian:2012mx}. Analogously,
in the $q{\bar q}$ annihilation channel
${\cal G}^{LL}(M,b)$ is the same as the corresponding contribution to the DY
process \cite{Catani:2015vma}.}
$gg \to t{\bar t}+X$ and $q{\bar q} \to t{\bar t}+X$.
By contrast, ${\cal G}^{NLL}(M,b)$ includes an important amount of
process-dependent information. In particular, it includes the effect of soft
wide-angle radiation for the $t{\bar t}$ production process.
This effect is controlled by a colour-dependent anomalous dimension, namely,
the soft anomalous dimension $\bf \Gamma^{(1)}_t$ in Ref.~\cite{Catani:2014qha}
(see Eqs.~(16), (17) and (33) in Ref.~\cite{Catani:2014qha}),
and we resummed the effect in complete exponentiated form throughout the 
diagonalization
of $\bf \Gamma^{(1)}_t$ in colour space.

At the technical level, we note that we compute 
$\Sigma^{(\rm res)}(M,{\bf b})$ 
in two-dimensional ${\bf b}$ space and we
obtain the $\qt$ cross section by numerically performing the Fourier
transformation to $\bqt$ space. The computation of the $\cos(n \varphi)$ harmonic
is also performed by numerically integrating over $\varphi$.
Therefore, we do not perform the projection onto
the $n$-th harmonic in analytic form
(see Eq.~(\ref{resn})),
and our implementation can be directly applied to the numerical evaluation of 
$n$-th harmonics with different values of $n$.

Our quantitative resummed results
on the $\qt$ dependence, $d\sigma/d\qt$,
of the $\cos(2\varphi)$ harmonic are presented in Fig.~\ref{tt2}-right.
We show the purely resummed result ($d\sigma^{(\rm res)}$) and the complete final
result after the inclusion of the matching term ($d\sigma^{(\rm fin)}$) at NLO.
In Fig.~\ref{tt2}-right we also report the NLO result (see Fig.~\ref{tt2}-left)
for direct comparison with the resummed results. The error bars in the complete
matched result denote the numerical errors of our calculation (for simplicity of
presentation the numerical error bars are not reported in the other histograms
of Fig.~\ref{tt2}). By direct inspection, we observe that the purely resummed
result quantitatively differs from the complete matched result, and we comment
about that below. Since we are considering the results after integration over the
invariant mass $M$ of the $t{\bar t}$ pair, the characteristic hard scale of the
$\qt$ cross section is of the order of $2m_t$ ($m_t$ is the mass of the top
quark), which is the minimum value of $M$,
 and the $\qt$ range that is shown in Fig.~\ref{tt2}-right includes the
regions of small and intermediate values of $\qt$.
The high-$\qt$ region, $\qt \gtap 2m_t$, is not shown in Fig.~\ref{tt2}-right.

We first comment on the results in the large-$\qt$ region of 
Fig.~\ref{tt2}-right. We observe that, starting from 
intermediate values of $\qt$ (e.g., $\qt \gtap 150$~GeV), the complete resummed
result tends to nicely agree with the NLO result. Note that we are not using a
direct (though, possibly smooth) switching procedure from the purely resummed
result to the NLO result at some intermediate value of $\qt$. The agreement
between the complete result and the NLO result at large $\qt$ is thus a
non-trivial consequence of the matching procedure. In the absence of the
contribution from the matching term  ($d\sigma^{(\rm fin)}$), the purely resummed
result sizeably overshoots the NLO result at intermediate and large values of
$\qt$ (for instance, at $\qt \sim 350$~GeV the purely resummed result is about
8 times larger than the NLO result). This feature is due to the fact that the
size of $d\sigma^{(\rm res)}$ is driven by the small-$\qt$ singular part of the
NLO calculation, and the singular part is not a good quantitative approximation
of the NLO calculation at intermediate and large values of $\qt$
(we have already discussed that and the corresponding overshooting effect
in our comments about Fig.~\ref{tt2}-left).
In the complete (matched) result, the effect of this approximation is
compensated by adding $d\sigma^{(\rm fin)}$ (i.e., the `NLO~$-$~sing' result
of Fig.~\ref{tt2}-left), which sizeably decreases $d\sigma^{(\rm res)}$
at large $\qt$. Once $d\sigma^{(\rm res)}$ and $d\sigma^{(\rm fin)}$
are combined, the difference between the resummed and NLO results 
is much reduced at intermediate values of $\qt$, and the higher-order (i.e.,
beyond NLO) resummation terms that are included in our resummed calculation
do not have a large residual effect in that $\qt$ region.

We now comment on the results in the small-$\qt$ region of 
Fig.~\ref{tt2}-right. In this region the resummed results unavoidably 
differ from the NLO result: the NLO result is much larger at small values of
$\qt$, and it diverges to $+\infty$ if $\qt \to 0$.
According to our previous discussion (see Eq.~(\ref{avqtn})), the purely resummed
result behaves as $d\sigma/d\qt \sim \qt^3$ in the limit $\qt \to 0$,
and the numerical results in Fig.~\ref{tt2}-right are consistent with this
expectation. The maximum value of the resummed cross section is in the region
($\qt$ bin) where 30~GeV$ < \qt < $60~GeV. Using our implementation of the
theoretical results of Ref.~\cite{Catani:2014qha}, we have performed a
corresponding resummed calculation of the customary (azimuthally-integrated) 
$\qt$ cross section $d\sigma/d\qt$ of the $t{\bar t}$ pair and we find
\cite{hayk} that its $\qt$ shape is quite different from that of the $\cos(2
\varphi)$ harmonic in Fig.~\ref{tt2}-right (as expected from the general comments
below Eq.~(\ref{avqtn})): $d\sigma_{\rm az. av.}/d\qt$ has a maximum at $\qt \sim
15$~GeV and a much softer $\qt$ spectrum in the small-$\qt$ region.
The qualitative features of the complete result for the 2nd harmonic are similar
to those of the corresponding purely resummed result.
However, the inclusion of the non-singular NLO contribution 
($d\sigma^{(\rm fin)}$) has a relevant quantitative effect, not only at
intermediate values of $\qt$ (as previously discussed) but also in the
small-$\qt$ region (for instance, in the $\qt$ bin where 
30~GeV$ < \qt < $60~GeV, the purely resummed result turns out to be reduced
by about 40\%). We have also checked the quantitative effect of the NLL
resummation terms that are due to the inclusion of 
the soft anomalous dimension $\bf \Gamma^{(1)}_t$. We find that also 
$\bf \Gamma^{(1)}_t$ has a substantial effect in the small-$\qt$ region:
$d\sigma/d\qt$ would be quantitatively smaller and with a harder $\qt$ shape
by removing the effect of $\bf \Gamma^{(1)}_t$ (i.e., by simply setting 
${\bf \Gamma^{(1)}_t} =0$).

By integrating $d\sigma/d\qt$ for the $\cos(2\varphi)$ harmonic over the entire
kinematical region of $\qt$, we find the total value 
$\sigma_{t{\bar t}, n=2}=3.0$~pb. In comparison, the value of the $t{\bar t}$
total cross section ($\sigma_{t{\bar t}}^{NLO}=226$~pb) is about 70 times larger
than $\sigma_{t{\bar t}, n=2}$.
We remark that our quantitative results for $\sigma_{t{\bar t}, n=2}$
and also for the $\qt$ dependence of the $\cos(2\varphi)$ harmonic
have to be regarded as `effective' lowest-order predictions within (resummed)
perturbative QCD. We do not attempt to quantify the theoretical accuracy of 
these predictions. As is well known, customary procedures (e.g., studies of
dependence on factorization and renormalization scales) that can be applied to
estimate theoretical uncertainties of {\em lowest-order} predictions typically
fail in reproducing the quantitative effect of higher-order contributions
(this is especially true in the case of QCD calculations for hard-scattering
processes in hadron collisions at high energies). A first quantitative estimate
of the theoretical uncertainty of the lowest-order predictions
requires a computation at the subsequent
perturbative order. Once the results at two subsequent orders are known, their
relative difference can be used for a quantitative estimate of the theoretical
uncertainty. At the present lowest-order level, we limit ourselves to adding 
few comments. The bulk of the contribution to the $\cos(2\varphi)$ harmonic
originates from the small and intermediate regions of $\qt$; we find that the
high-$\qt$ region (say, $\qt \gtap 2m_t$) gives only a few percent contribution
to the total value of $\sigma_{t{\bar t}, n=2}$.  The NLL resummed contributions
due to the soft anomalous dimension $\bf \Gamma^{(1)}_t$ are not negligible:
removing the effect due to $\bf \Gamma^{(1)}_t$ in our calculation, we find that 
the total value of $\sigma_{t{\bar t}, n=2}$ decreases by about 30\%.
The total value of $\sigma_{t{\bar t}, n=2}$ would be much larger by considering
the purely resummed term ($d\sigma^{(\rm res)}$) and no matching with the
complete NLO calculation: removing the matching contribution 
($d\sigma^{(\rm fin)}$), the value of  $\sigma_{t{\bar t}, n=2}$
would increase by roughly a factor of two. This highlights the importance of
non-singular NLO effects at small and intermediate values of $\qt$.

In our resummed calculation we numerically perform the Fourier transformation
from $\bf b$-space to $\bqt$-space, and we obtain the $\qt$ cross section
$d\sigma/d\qt^2 d\varphi$. We can then numerically evaluate different harmonics.
We have carried out an exploratory numerical computation of the $\cos(4\varphi)$
asymmetry for $t{\bar t}$ production, and we find results with the expected
features, analogously to the $\cos(2\varphi)$ asymmetry. Nonetheless, the size of the
$\cos(4\varphi)$ asymmetry is very small and we do not present corresponding
resummed results. As shown in Fig.~\ref{dyfig}-right, 
the $n=1$ azimuthal asymmetry for
$t{\bar t}$ production is not divergent if computed at its lowest perturbative
order,
and the corresponding $\qt$ spectrum is not singular. In view of our discussion
in Sect.~\ref{sec:azor}, the next-order QCD correction to the  
$n=1$ harmonic can produce a singular $\qt$ spectrum, so that the corresponding 
QCD prediction can require a resummed calculation at a subdominant logarithmic
accuracy with respect to that considered in Ref.~\cite{Catani:2014qha}.

\setcounter{footnote}{2}

We add some general comments on the contribution of non-singular terms
(e.g., the terms denoted by dots in the right-hand side of Eq.~(\ref{nlosigma}))
to azimuthal correlations at low values of $\qt$.
In the case of the azimuthally-averaged cross section
$d\sigma_{\rm az. av.}/d\qt^2$, the perturbative resummation of singular terms
produces a constant behaviour in the small-$\qt$ region
(see Eq.~(\ref{avqt0})) and, typically (e.g., in absence of additional
kinematical cuts), the non-singular terms also have a constant behaviour at low
values of $\qt$ (order-by-order in the perturbative expansion the ratio between
non-singular and singular terms if formally of ${\cal O}(\qt^2/M^2)$, modulo
logarithms, in the limit $\qt \to 0$).
As a consequence, resummed and f.o. non-singular contributions to 
$d\sigma_{\rm az. av.}/d\qt^2$ behave similarly at low values of $\qt$,
although the non-singular
contributions are perturbatively (and quantitatively) suppressed.
In the case of the cross section $d\sigma_n/d\qt^2$ for the $n$-th harmonic,
the resummation of the singular terms produces an enhanced power suppression
($\propto \qt^n$) of dynamical origin (see Eq.~(\ref{avqtn}))
in the small-$\qt$ region and, therefore,
non-singular terms are eventually more relevant at low values of $\qt$ if they
are treated at f.o. in perturbation theory. 
We note that this reasoning neglects the fact that non-singular terms can also
have logarithmic enhancement (powers of $\ln(M/\qt)$) order-by-order in
perturbation theory and, consequently, an appropriate resummation treatment of
non-singular terms can be required. For instance, QCD resummation of non-singular
terms has been investigated 
(see Ref.~\cite{Berger:2007jw} and references therein)
for the DY process. Note, however, that non-singular corrections to singular
azimuthal correlations are quite different from azimuthal-correlation effects in
the DY process\footnote{For instance, in a previous footnote we have already
remarked that in the small-$\qt$ region the 2nd harmonic cross section
$d\sigma_{n=2}/d\qt^2$ has non-singular terms of ${\cal O}(M/\qt)$ for
$t{\bar t}$ production and of ${\cal O}((\qt/M)^0)= {\cal O}(1)$ for the DY
process.}
and they are expected to have a high degree of process dependence.
We also recall that the non-singular terms produce
azimuthal-correlation effects whose actual size and $\qt$ behaviour depend
on the specific definition of the azimuthal-correlation angles (see 
Eq.~(\ref{angles}) and related comments in Sect.~\ref{sec:corr}). 
In particular, the $\qt$ dependence of non-singular terms can be redistributed
among different $n$-th harmonics by considering azimuthal correlations that refer
to different specifications of the azimuthal angle.

As for non-singular terms, 
at present we do not have a detailed theoretical understanding 
that goes beyond their treatment at f.o.\,.
A better understanding of non-singular terms is certainly relevant to have an
accurate quantitative control on the detailed $\qt$ shape of azimuthal
correlations in the region of very-low values of $\qt$. We note, however,
that non-singular terms have a relatively-mild effect on $\qt$ integrated
azimuthal correlations. For instance, since non-singular terms are {\em
integrable} in the limit $\qt \to 0$, their perturbative logarithmic enhancement
still gives contributions to the total ($\qt$ integrated) harmonic that are
parametrically controlled by powers\footnote{This effect is analogous to the
corresponding effect on the total ($\qt$ integrated) azimuthally-averaged cross
section: its value is reliably computable at f.o. independently of the presence
of {\em integrable} logarithmically-enhanced, and even singular
(plus-distributions), contributions in the small-$\qt$ region.} of $\as$.
Moreover, after resummation of the singular terms (and, basically, 
as a consequence
of the behaviour in Eq.~(\ref{avqtn})),
the bulk of the contribution to $d\sigma_n/d\qt^2$
is located at relatively-large values of $\qt$ (the bulk of the contribution to 
$d\sigma_{\rm az. av.}/d\qt^2$ is typically located at smaller values of $\qt$)
and, also in this region the logarithmic enhancement of non-singular terms is
expected to produce relatively-mild effects (approximately, also these effects are
parametrically controlled by powers of $\as$).
In summary, we think that an improved (beyond f.o. perturbation theory)
treatment of non-singular terms is relevant especially to determine the {\em
detailed} (point-by point) $\qt$ shape of $d\sigma_n/d\qt^2$ at low values of
$\qt$, while even the total effect of non-singular terms in the low-$\qt$ region
(e.g., in the lower $\qt$-bin, $\qt < 30$~GeV, of Fig.~\ref{tt2})
can be relatively well under control within a f.o. treatment.

We note that, at very low values of $\qt$, also non-perturbative QCD effects can
be relevant. In the case of the resummed contribution to the $\qt$ cross section,
a customary (though much simplified) procedure to model non-perturbative effects
consists in supplementing the Sudakov form factor with a non-perturbative form
factor that mostly contributes at large values of $b$ in the $b$-space resummation
formula (see Eq.~(\ref{resav})). A typical parametrization of the 
non-perturbative form factor is $e^{-g\, b^2}$ ($g$ being a parameter whose size,
$g \sim {\cal O}(1\!~{\rm GeV}^2)$, is determined by effects at non-perturbative
scales), and it produces a strong damping effect a very large values of $b$. 
In the resummation formula these
non-perturbative effects act in a $b$-region that is already strongly suppressed
by the Sudakov form factor and, consequently (after Bessel transformation from $b$
space to $\qt$ space), they mainly affect only the region of low values of $\qt$
\cite{Parisi:1979se, Collins:va}.
Such a model of non-perturbative effects can also be applied to the resummed
contribution to the $n$-th harmonic (see Eq.~(\ref{resn})), and the 
non-perturbative parameter can also depend on the harmonic (e.g., $g \to g_n$).
We may try to apply a related model also to non-singular terms.
We consider (part of) non-singular terms of the $n$-th harmonic, we transform them
to $b$-space and then we supplement them with a non-perturbative form factor
$e^{-g_n\, b^2}$. The non-singular terms in $b$-space are power-suppressed at
large values of $b$, so that the non-perturbative form factor
still acts on a $b$-region that is dynamically suppressed
(correspondingly, the non-perturbative form factor mainly affects the region of
low values of $\qt$). Nonetheless, at very large values of $b$, the exponential
suppression that is produced by the non-perturbative form factor
is so strong that, in the limit $\qt \to 0$, $d\sigma_n/d\qt^2$ 
behaves\footnote{We are simply applying the same reasoning that leads to
Eq.~(\ref{avqtn}), by replacing the role of the Sudakov form factor with that of
the non-perturbative form factor.}
as $\qt^n$, and the non-singular terms (those that are treated in this way)
eventually produce the same $\qt$ behaviour as the resummed contribution to
$d\sigma_n/d\qt^2$. We cannot argue that such a model is physically justified
(such a non-perturbative treatment of non-singular terms is certainly not
applicable to azimuthal correlations in a pure QED context). We have mentioned it
only to notice that in the region of low values of $\qt$ non-singular terms can be
affected by both perturbative logarithmic enhancement and non-perturbative
contributions. A study of non-perturbative effects in azimuthal-correlation cross
sections is definitely beyond the scope of this paper.

\section{Summary}
\label{sec:summa}

In this paper we have considered high-mass systems
formed by two or more particles that are produced in hadron collisions,
and we have discussed the azimuthal correlations between the system and one 
of its particles.
We refer to particles in an extended sense, namely, pointlike particles and 
QCD jets. Our main findings can be summarized as follows.

Despite the infrared-safe nature of the azimuthal distribution, the f.o.
QCD computation of this observable can be divergent
starting from some perturbative order.
This conclusion holds 
for a large class of processes, including $t\bar t$, $Vj$, $jj$, 
$\gamma\gamma$, $ZZ$, $W^+W^-$ production, while the corresponding QCD 
calculation for the DY process is finite to any (arbitrary) perturbative order.
We have supported this conclusion by performing QCD calculations at 
the first non-trivial order
for the DY, $Z$+jet and $t{\bar t}$ production processes.

The divergence arises from a complete mismatch between virtual and real 
contributions at fixed perturbative orders.
More precisely, the divergence in the real contribution is not compensated 
by a corresponding virtual term, since the latter cannot give azimuthal
correlations. This implies that quantities such as
$d\sigma/dM^2dq_T^2d\varphi$ are singular in the limit $q_T\to 0$ (they are
proportional to $1/\qt^2$, modulo powers of $\ln(M/\qt)$)
and, more
importantly, their total integral over $q_T^2$ is divergent.
Similar features and divergences occur by considering
QED radiative corrections to the DY process.

The origin of the divergences in the azimuthal correlations can be identified in
the existence of singular contributions from different sources. If one of the
initial-state colliding partons is a gluon, spin correlations of collinear 
origin produce
singular contributions in specific azimuthal harmonics, namely, harmonics with
$n=2$ and $n=4$. If at
least one of the triggered final-state particles carries QCD colour charge, 
soft-gluon radiation at large angles produces singular contributions that 
in principle affect harmonics with {\it any} value of $n$.
Divergent QED corrections are produced by soft-photon radiation 
if least one of the triggered final-state particles carries 
non-vanishing electric charge.

The f.o. divergences in the azimuthal correlations can be tamed by all-order 
resummation of the singular contributions.
The resummation procedure is carried out by working in ${\bf b}$-space, 
analogously to what is done in the customary transverse-momentum resummation
of azimuthally-averaged (or azimuthally-insensitive) cross sections
$d\sigma/dM^2dq_T^2$.
We have discussed the general behavior of the resummed azimuthal-correlation 
cross section in the small-$\qt$ region
by contrasting it with the case of the azimuthally-averaged 
cross section.
Owing to the dynamical suppression of the large-$b$ region that is produced
by the (resummed) Sudakov form factor,
in the azimuthally-averaged case  
resummation transforms the $1/q_T^2$ 
singular behavior at f.o. into a constant behavior in the region of small values
of $q_T$.
In the case of 
azimuthal asymmetries, the effect of resummation is even more substantial, 
since the $1/q_T^2$ behavior of the $n$-th harmonic at f.o.
is turned into a power-like behavior proportional to $q_T^n$.
More importantly, in the case of the azimuthally-correlated cross section, 
resummation allows us to obtain a prediction for the total ($\qt$ integrated)
cross section $d\sigma/dM^2d\varphi$.
This should be contrasted with the case of the azimuthally-averaged inclusive 
cross section $d\sigma/dM^2$, which is finite and computable in f.o. 
perturbation theory and, hence, it does not require resummation.

The $t{\bar t}$ production process can be chosen as a prototype to present 
quantitative results.  
Besides its obvious phenomenological relevance,
the main reason to focus our attention on this process is that, 
due to its gluon-fusion production channel and to 
the non-vanishing colour charges of the produced $t$ and ${\bar t}$, 
the process features 
singular azimuthal-correlation effects of both collinear and soft origin.
We have considered the $\cos(2\varphi)$ harmonic and
we have presented numerical results at NLL+NLO accuracy
for $t{\bar t}$ production in $pp$ collisions at the LHC.
The resummed $q_T$ spectrum of the $n=2$ harmonic is finite as $q_T\to 0$, 
and it displays
a Sudakov peak that is located at higher $q_T$ values with respect to the peak 
of the azimuthally-averaged distribution.
This is consistent with the expectations from resummation of azimuthal
correlations.
The integral over $\qt$ of the resummed distribution 
leads to an `effective' lowest-order prediction for the total value of
the $n=2$ harmonic. More detailed results will be presented elsewhere.

The discussion and the results presented in this paper have a high generality. 
They set the stage for obtaining
perturbative predictions for azimuthal asymmetries in a wide class of processes,
in which f.o. perturbation theory returns divergent, and thus useless, results. 
Besides heavy-quark production, we envisage interesting applications to 
vector boson plus jet and dijet production at hadron colliders,
and we anticipate further applications
to hard-scattering processes in $ep$ and $e^+e^-$ collisions.

\noindent {\bf Acknowledgements}. We would like to thank
Alessandro Bacchetta for discussions.
This research was supported in part by the Swiss National Foundation (SNF)
under contract 200020-169041, by the Forschungskredit of the University of
Zurich and by the Research Executive Agency (REA) of the European Union under
the Grant Agreement number PITN-GA-2012-316704 ({\em Higgstools}).

\end{document}